\documentclass[aps,showkeys,twocolumn,superscriptaddress,floatfix,letterpaper]{revtex4}

\usepackage{mathtools}
\usepackage{epstopdf}
\usepackage{braket}
\usepackage{amssymb}
\usepackage{simplewick}

\begin{document}

\title{Nonlinear Power Spectral Densities for the Harmonic Oscillator}

\author{B. D. Hauer}
\email{bhauer@ualberta.ca}
\address{Department of Physics, University of Alberta, Edmonton, AB, Canada T6G 2G7}
\author{J. Maciejko}
\address{Department of Physics, University of Alberta, Edmonton, AB, Canada T6G 2G7}
\address{Theoretical Physics Institute, University of Alberta, Edmonton, AB, Canada T6G 2G7}
\address{Canadian Institute for Advanced Research, Toronto, ON, Canada M5G 1Z8}
\author{J. P. Davis}
\email{jdavis@ualberta.ca}
\address{Department of Physics, University of Alberta, Edmonton, AB, Canada T6G 2G7}

\date{\today}

\begin{abstract}

In this paper, we discuss a general procedure by which nonlinear power spectral densities (PSDs) of the harmonic oscillator can be calculated in both the quantum and classical regimes. We begin with an introduction of the damped and undamped classical harmonic oscillator, followed by an overview of the quantum mechanical description of this system. A brief review of both the classical and quantum autocorrelation functions (ACFs) and PSDs follow. We then introduce a general method by which the $k$th-order PSD for the harmonic oscillator can be calculated, where $k$ is any positive integer. This formulation is verified by first reproducing the known results for the $k=1$ case of the linear PSD. It is then extended to calculate the second-order PSD, useful in the field of quantum measurement, corresponding to the $k=2$ case of the generalized method. In this process, damping is included into each of the quantum linear and quadratic PSDs, producing realistic models for the PSDs found in experiment. These quantum PSDs are shown to obey the correspondence principle by matching with what was calculated for their classical counterparts in the high temperature, high-$Q$ limit. Finally, we demonstrate that our results can be reproduced using the fluctuation-dissipation theorem, providing an independent check of our resultant PSDs.

\end{abstract}


\maketitle

\section{Introduction}
\label{intro}

The harmonic oscillator, in which a particle is confined to a potential well that varies quadratically with position, has proven to be a very useful model in a number of classical and quantum systems. In the classical regime, the harmonic oscillator provides an excellent description of periodic systems such as a mass on a spring or a pendulum, as well as resonating electronic LC circuits. In the realm of quantum mechanics, an analogous model is successful in predicting the behavior of a number of bosonic systems, such as photons confined to an optical cavity or phonons in an elastic solid. In fact, the vacuum itself is thought to consist of an array of harmonic oscillators with a broad range of frequencies \cite{gardiner}. 

Often, a harmonic oscillator model is applied to a system in isolation, where we generally consider only linear effects.  However, when we begin to consider coupling between harmonic oscillators, or with other systems altogether, nonlinearities begin to enter the model, leading to new physics. An example of this sort of interaction arises in cavity optomechanics, in which two harmonic oscillators, one describing an optical cavity and the other describing a mechanical resonator, are coupled to one another \cite{aspelmeyer}. In this case, the motion of the mechanical resonator shifts the resonance frequency of the optical cavity, while the optics provide a radiation pressure force acting back on the mechanics. For moderate coupling, a simple linear model suffices, such that monitoring the electromagnetic field provides a readout of the linear motion of the oscillating mechanical device. However, as the interaction strength between the two systems increases, nonlinear coupling begins to occur, requiring that higher-order terms be added to the Hamiltonian \cite{thompson,sankey,nunnenkamp,gangat,huang,borkje,doolin}. This provides a method by which one can obtain direct access to higher-order powers of the mechanical resonator's motion. For instance, a number of experiments have demonstrated direct coupling to the square of the oscillator's displacement \cite{thompson,sankey,doolin,purdy,hill}. These types of measurements have generated significant interest, as they have been proposed as a method to perform quantum nondemolition (QND) measurements \cite{braginsky,braginsky2} of a mesoscopic quantum system \cite{aspelmeyer,thompson,gangat,jayich,chen,clerk1,kaviani}, as well as other exotic two-phonon processes, such as mechanical cooling/squeezing \cite{nunnenkamp} and optomechanically induced transparency \cite{huang,borkje}. 

In order to make such measurements effectively, a knowledge of the autocorrelation functions (ACFs) and power spectral densities (PSDs) corresponding to the nonlinear readout of the oscillator's motion is required. Though the first-order PSD is a well-known result  \cite{hauer,albrecht,ekinci,krause,clerk2,safavi-naeini2}, here we calculate a general PSD of any order for the quantum and classical harmonic oscillator, with a special focus on the linear and quadratic cases. The structure of this document is as follows. In Section \ref{back}, we provide a basic overview of the classical and quantum harmonic oscillators in the damped and undamped situations. Section \ref{ACFPSD} then provides a description of how to calculate the ACF and PSD of a classical time-dependent signal. Complementary definitions for a time-dependent quantum operator follow. Using the results of the previous two sections, Section \ref{PSDk} introduces a general procedure that can be used to calculate the classical and quantum PSDs of $k$th-order for the harmonic oscillator. Section \ref{PSD1} reviews the $k=1$ case of the first-order PSD of the harmonic oscillator, which is immediately followed by an extension to the $k=2$ case of the quadratic PSD in Section \ref{PSD2}. Finally, we conclude the document by discussing how these PSDs can be used in the context of real experiments.

\section{Background}
\label{back}

\subsection{Classical Undamped Harmonic Oscillator}
\label{CHO}

The model of the classical, undamped harmonic oscillator describes a system whose dynamics are governed by the following differential equation
\begin{equation}
\ddot{x} + \omega_0^2 x = 0,
\label{CHOdiff}
\end{equation}
\noindent where $x(t)$ is a time-dependent variable that in this case we choose to be the position of the oscillator and $\omega_0 = \sqrt{k /m}$ is the resonant angular frequency of the system, with $k$ and $m$ being the oscillator's spring constant and mass, respectively. The familiar oscillatory solution to this second-order differential equation is given by 
\begin{equation}
x(t) = x_0 \cos (\omega_0 t + \phi),
\label{CHOsol}
\end{equation}
\noindent where $x_0$ and $\phi$ are an arbitrary amplitude and phase of the motion set by the initial conditions. 

We can determine the total energy of this system as the sum of its kinetic and potential energies. The potential for this system is $V = \frac{1}{2} k x^2 = \frac{1}{2} m \omega_0^2 x^2$, while the kinetic energy is simply the conventional $K = \frac{1}{2} m \dot{x}^2 = \frac{p^2}{2m}$, where $p$ is the linear momentum of the one-dimensional system. Using our solution for $x(t)$ from above, we find the total energy to be
\begin{equation}
\begin{split}
E &= H = K + V = \frac{p^2}{2m} + \frac{1}{2} m \omega_0^2 x^2 \\
&= \frac{m \omega_0^2 x_0^2 }{2} \left[\sin^2 (\omega_0 t + \phi) + \cos^2 (\omega_0 t + \phi) \right] \\
&= \frac{m \omega_0^2 x_0^2 }{2},
\end{split}
\label{CHOenergy}
\end{equation}
\noindent which is a time-independent quantity. Note that in this case, we can equate the total energy to the Hamiltonian of the system, which we have denoted as $H$.

\subsection{Classical Damped Harmonic Oscillator}
\label{CDHO}

While the undamped harmonic oscillator provides the simplest solution to an oscillatory problem, this model can be made more realistic by introducing damping into the system, allowing for the description of real-world dissipative systems, including LRC circuits and nanomechanical resonators \cite{hauer}. The most straightforward way to introduce damping into Eq.~(\ref{CHOdiff}) is to add a term proportional to $\dot{x}(t)$, producing the new differential equation
\begin{equation}
\ddot{x} + \Gamma \dot{x} + \omega_0^2 x = 0,
\label{CDHOdiff}
\end{equation}
\noindent where $\Gamma$ is a characteristic rate that quantifies the damping in the system.

In the underdamped case $\left( \Gamma < 2 \omega_0 \right)$, the solution to this equation is given by
\begin{equation}
x(t) = x_0 e^{-\frac{\Gamma t}{2}} \cos (\omega_d t + \phi),
\label{CDHOsol}
\end{equation}
\noindent where $\omega_d = \omega_0 \sqrt{1 - \left( \Gamma / 2 \omega_0 \right)^2}$ is the shifted resonance frequency due to damping. When damping is very small $\left( \Gamma \ll 2 \omega_0 \right)$, we neglect this shift and take $\omega_d \approx \omega_0$.  In this limit, Eq.~(\ref{CHOsol}) provides a good approximation for the solution of the damped harmonic oscillator given by Eq.~(\ref{CDHOsol}). From this point forward, we will assume we are in the small damping limit, as this is the case of interest for most nanomechanical systems.

Another useful parameter which can be used to quantify the damping of the system described above is the quality factor $Q$, defined by the equation
\begin{equation}
Q = 2 \pi \frac{E}{\Delta E},
\label{Qdef}
\end{equation}
\noindent where $\Delta E$ is the energy dissipated per oscillation cycle. For the damped harmonic oscillator given above, we calculate the total energy of the system as we did in the undamped case, resulting in
\begin{equation}
\begin{split}
E &= \frac{m \omega_0^2 x_0^2 }{2} e^{-\Gamma t} \left[\sin^2 (\omega_0 t + \phi) + \cos^2 (\omega_0 t + \phi) \right]  \\
&= \frac{m \omega_0^2 x_0^2 }{2}  e^{-\Gamma t},
\end{split}
\label{CDHOenergy}
\end{equation}
\noindent where in the above equation we have neglected a term in $\dot{x}(t)$ that is proportional to $\Gamma$ as we are in the small damping limit. The result for the energy of the damped harmonic oscillator is identical to that for its undamped counterpart given in Eq.~(\ref{CHOenergy}), except now the energy decays on a timescale set by $\Gamma$. The energy dissipated in one cycle is then given by the change of energy over one period of oscillation $\tau_0$, that is
\begin{equation}
\begin{split}
\Delta E &= \frac{m \omega_0^2 x_0^2 }{2}  e^{-\Gamma t} - \frac{m \omega_0^2 x_0^2 }{2}  e^{-\Gamma (t + \tau_0)} \\
&= \frac{m \omega_0^2 x_0^2 }{2}  e^{-\Gamma t} \left( 1 - e^{- \Gamma \tau_0} \right).
\end{split}
\label{CDHOdelenergy}
\end{equation}
\noindent The quality factor for this system is then given by
\begin{equation}
Q = \frac{2 \pi}{1 - e^{-\Gamma \tau_0}} \approx \frac{2 \pi}{1 - \left( 1 - \Gamma \tau_0 \right)} = \frac{\omega_0}{\Gamma},
\label{Qdamp}
\end{equation}
\noindent where we have again used the small damping limit and the fact that we can relate the period of oscillation to the system's angular resonant frequency via $\tau_0 = 2 \pi / \omega_0$. From Eq.~(\ref{Qdamp}) it becomes apparent that the small damping limit is equivalent to the high-$Q$ limit, as smaller damping leads to a reduction in energy dissipation. In fact, using our condition for the small damping limit above, we can quantify the high-$Q$ limit as $Q \gg 1/2$. For the remainder of the document, we will refer to the small damping limit as the high-$Q$ limit.

By analyzing the undriven, damped harmonic oscillator above, we were able to investigate the time domain solution of the resonator's motion, as well as its energy dissipation. However, this description is still somewhat incomplete as generally the motion will be driven by some time-dependent external driving force $f(t)$. In such a situation, we arrive at the driven differential equation of motion
\begin{equation}
\ddot{x} + \Gamma \dot{x} + \omega_0^2 x = \frac{f}{m}.
\label{CDHOdrivediff}
\end{equation}
\noindent Analytical solutions for $x(t)$ in this case can only be determined for a small number of special cases of $f(t)$, such as a harmonic driving force. However, it is often more fruitful to Fourier transform this equation to get its expression in the frequency domain, resulting in
\begin{equation}
x(\omega) = \chi(\omega) f(\omega) = \frac{ f(\omega)}{m \left( \omega_0^2 - \omega^2 - i \omega \Gamma \right)},
\label{CDHOfreq}
\end{equation}
\noindent where $x(\omega) = \mathcal{F} \{ x(t) \}$ and $f(\omega) = \mathcal{F} \{ f(t) \}$ are the Fourier transforms of $x(t)$ and $f(t)$ as defined in \ref{four} and we have used the property in Eq.~(\ref{fourderiv}) to calculate the Fourier transforms of the derivatives. We have also introduced the generalized mechanical susceptibility
\begin{equation}
\chi(\omega) = \frac{1}{m \left( \omega_0^2 - \omega^2 - i \omega \Gamma \right)},
\label{mechchi}
\end{equation}
\noindent which allows us to relate the resultant position to the applied force in the frequency domain.

\subsection{Quantum Harmonic Oscillator}
\label{QHO}

To extend the above treatment of the harmonic oscillator into the quantum domain, we must first determine its governing quantum mechanical Hamiltonian. This is accomplished by simply replacing $x$ and $p$ in the first line of Eq.~(\ref{CHOenergy}) with the canonically conjugate position and momentum operators $\hat{x}$ and $\hat{p}$. With these new operators, our kinetic and potential energies now become $\hat{K} = \frac{\hat{p}^2}{2m}$ and $\hat{V} = \frac{1}{2} m \omega_0^2 \hat{x}^2$, resulting in
\begin{equation}
\hat{H} = \hat{K} + \hat{V} = \frac{\hat{p}^2}{2m} + \frac{1}{2} m \omega_0^2 \hat{x}^2.
\label{QHOham}
\end{equation}
\noindent Inputting this Hamiltonian into the Schr\"{o}dinger equation, it is possible to solve for the energy eigenstates of this system $\ket{n}$, along with their corresponding energy eigenvalues $E_n$, where $\hat{H} \ket{n} = E_n \ket{n}$. For this derivation, we do not concern ourselves with the exact form of the eigenstates, however, the energies are given by
\begin{equation}
 E_n = \hbar \omega_0 \left( n + \frac{1}{2} \right),
\label{QHOenergy}
\end{equation}
\noindent where $\hbar = h / 2 \pi$ is the reduced Planck's constant. In the above equation, $n$ is an integer and signifies the state of the oscillator. Quantum mechanically, this number $n$ can be interpreted as the number of quanta in the system, for example photons in a cavity or phonons in a solid. Therefore, $E_0 = \frac{\hbar \omega_0}{2}$ denotes the ground state energy where $n=0$ and no quanta exist in the system. This energy will be shared evenly between the expectation values of the kinetic and potential energy such that $\langle \hat{K} \rangle = \langle \hat{V} \rangle = \frac{\hbar \omega_0}{4}$ in the ground state.

We now introduce the raising (creation) and lowering (annihiliation) operators $\hat{b}^{\dag}$ and $\hat{b}$, also known as the ladder operators. These two quantities are given by
\begin{equation}
\begin{split}
\hat{b} &= \sqrt{ \frac{m \omega_0}{2 \hbar}}\left( \hat{x} + \frac{i}{m \omega_0} \hat{p} \right), \\
\hat{b}^{\dag} &= \sqrt{ \frac{m \omega_0}{2 \hbar}} \left( \hat{x} - \frac{i}{m \omega_0} \hat{p} \right),
\label{bops}
\end{split}
\end{equation}
\noindent and obey the commutation relation $[\hat{b} , \hat{b}^{\dag}] = 1$. These operators are convenient as they produce the following simple relations when operating on the energy eigenstates of the system
\begin{equation}
\begin{split}
\hat{b} \ket{n} &= \sqrt{n} \ket{n-1}, \\
\hat{b}^{\dag} \ket{n} &= \sqrt{n+1} \ket{n+1},
\label{raiselower}
\end{split}
\end{equation}
\noindent as well as their Hermitian conjugates
\begin{equation}
\begin{split}
\bra{n} \hat{b}^{\dag}  &= \bra{n-1} \sqrt{n}, \\
\bra{n} \hat{b} &= \bra{n+1} \sqrt{n+1}.
\label{raiselowerh}
\end{split}
\end{equation}
\noindent From these relations we also have
\begin{equation}
\begin{split}
\hat{b}^{\dag} \hat{b} \ket{n} &= n \ket{n}, \\
\hat{b} \hat{b}^{\dag} \ket{n} &= (n+1) \ket{n}.
\label{numops}
\end{split}
\end{equation}
\noindent As seen above, when acting on the energy eigenstates the operator $\hat{N} = \hat{b}^{\dag} \hat{b}$ returns the number of quanta $n$ of that state and is known as the number operator. By inspecting Eq.~(\ref{QHOenergy}) it should therefore be clear that the Hamiltonian can be expressed as
\begin{equation}
\hat{H} = \hbar \omega_0 \left( \hat{b}^{\dag} \hat{b} + \frac{1}{2} \right) = \hbar \omega_0 \left( \hat{N} + \frac{1}{2} \right).
\label{QHOhamnum}
\end{equation}
\noindent We can also write $\hat{x}$ in terms of the ladder operators as
\begin{equation}
\begin{split}
\hat{x} &= x_{\rm zpf} \left( \hat{b} + \hat{b}^{\dag} \right),
\label{xop}
\end{split}
\end{equation}
\noindent where we have introduced $x_{\rm zpf} = \sqrt{\frac{\hbar}{2 m \omega_0}}$, which is the amplitude of the quantum mechanical zero point fluctuations of the oscillator.

Up to this point, we have been dealing with operators in the Schr\"odinger picture, where it is the eigenstates, not the operators, that carry the time-dependence of the problem. However, since we are dealing with time-dependent signals, it is convenient to turn to the Heisenberg picture of quantum mechanics, where the operators are now the quantities that vary in time. The dynamics of an operator $\hat{O}$, which was time-independent in the Schr\"odinger picture, is now governed by the differential equation
\begin{equation}
\dot{\hat{O}} = \frac{i}{\hbar} [ \hat{H}, \hat{O} ].
\label{heis}
\end{equation}
\noindent Upon inspection of this equation, we see that an operator which is time-independent in the Schr\"odinger picture and commutes with the Hamiltonian will remain constant in the Heisenberg picture. 

Using the Hamiltonian for the quantum harmonic oscillator given in Eq.~(\ref{QHOhamnum}), along with the commutation relation for the ladder operators, we can obtain a differential equation for $\hat{b}(t)$ given by
\begin{equation}
\dot{\hat{b}} = -i \omega_0 \hat{b}.
\label{bdiff}
\end{equation}
\noindent This equation is easily integrated to obtain the expression for the annihilation and creation operators in the Heisenberg picture as
\begin{equation}
\begin{split}
\hat{b}(t) &= \hat{b}e^{-i \omega_0 t}, \\
\hat{b}^{\dag}(t) &= \hat{b}^{\dag}e^{i \omega_0 t},
\end{split}
\label{boft}
\end{equation}
\noindent where the latter equation is obtained by taking the adjoint of the former.

Finally, we determine a time-varying expression for $\hat{x}(t)$ in the Heisenberg picture by inputting the relations in Eq.~(\ref{boft}) into Eq.~(\ref{xop}) to obtain
\begin{equation}
\begin{split}
\hat{x}(t) &= x_{\rm zpf} \left( \hat{b}e^{-i \omega_0 t} + \hat{b}^{\dag}e^{i \omega_0 t} \right). \\
\label{xopt}
\end{split}
\end{equation}

Another advantage of working in the Heisenberg picture is that we can introduce damping into Eq.~(\ref{bdiff}) through a formalism known as input-output theory \cite{gardiner}. To do this, we assume that our harmonic oscillator is coupled to a bath which has some effective temperature $T$. Through this coupling, the oscillator is able to reach thermal equilibrium with the bath, by either losing energy to it or gaining energy from it, corresponding to damping of the oscillator and an incoherent drive from the bath. In general, this drive will have contributions originating from both the thermal occupation of the bath, as well as its quantum mechanical fluctuations. 

Often, this bath is chosen to be an ensemble of harmonic oscillators with varying resonance frequencies, all of which are at the bath temperature. This method proves to be very effective, as it is solvable due to the simplicity of the harmonic oscillator, and provides an accurate model of physically realizable baths, such an electromagnetic field or phonons in a solid \cite{gardiner}. Using this model, along with the first Markov approximation (memoryless coupling to the bath quantified by a constant) \cite{gardiner}, we modify Eq.~(\ref{bdiff}) to obtain a new equation of motion
\begin{equation}
\dot{\hat{b}}_{\gamma} = -i\omega_0 \hat{b}_{\gamma} - \frac{\Gamma}{2} \hat{b}_{\gamma} + \sqrt{\Gamma} \hat{b}_n,
\label{bdampdiff}
\end{equation}
\noindent where we have introduced a subscript $\gamma$ to differentiate this ladder operator from the undamped one. In the above equation, $\Gamma$ quantifies the coupling of our oscillator to the bath and corresponds directly to the mechanical damping rate mentioned above for the classical case. This is exemplified by the fact that if we set $\Gamma = 0$ in Eq.~(\ref{bdampdiff}), coupling to the bath is severed and we reclaim the original, undamped differential equation given by Eq.~(\ref{bdiff}). 

In the above equation, the two new terms have arisen from coupling our oscillator to the bath. The second term on the RHS describes a decay in the amplitude of $\hat{b}_\gamma(t)$ due to energy radiation to the bath, while the third term represents the drive due to input noise from the bath, given by the operator $\hat{b}_n(t)$. We assume that this noise will be delta-correlated  in time ({\it i.e.}~Markovian), which corresponds closely to classical white noise, resulting in \cite{gardiner,aspelmeyer}
\begin{equation}
\begin{split}
\braket{\hat{b}_n(t) \hat{b}^{\dag}_n(t') } &= \left( n_b + 1 \right) \delta(t-t'), \\
\braket{\hat{b}^{\dag}_n(t) \hat{b}_n(t') } &=  n_b  \delta(t-t'), \\
\braket{\hat{b}^{\dag}_n(t) \hat{b}^{\dag}_n(t') } &=  \braket{\hat{b}_n(t) \hat{b}_n(t') } = 0.
\label{noisecorrt}
\end{split}
\end{equation}
\noindent Assuming that the bath occupation will be constant over the small bandwidth $\sim \Gamma$ of interest about the oscillator's resonance frequency we can take the bath occupation number to be the single value $n_b = n_b(\omega_0)$. 

In this case, it is difficult to obtain a time-domain representation for $\hat{b}_{\gamma}(t)$ due to the noise input into the system. Instead, we Fourier transform Eq.~(\ref{bdampdiff}) to obtain the spectral form of the annihilation operator
\begin{equation}
\hat{b}_{\gamma} (\omega) = \frac{\sqrt{\Gamma} \hat{b}_n(\omega)}{i \left( \omega_0 - \omega \right) + \Gamma / 2},
\label{bofw}
\end{equation}
\noindent where we have introduced the Fourier transformed operators $\hat{b}_{\gamma}(\omega) = \mathcal{F} \{ \hat{b}_{\gamma}(t) \}$ and $\hat{b}_n(\omega) = \mathcal{F} \{ \hat{b}_n(t) \}$. We can also determine the spectral form of the creation operator, $\hat{b}^{\dag}_{\gamma}(\omega) = \mathcal{F} \{ \hat{b}^{\dag}_{\gamma}(t) \}$, by taking the adjoint of the above equation and using the relation $[ \hat{b}_{\gamma}(\omega)]^{\dag} = \hat{b}_{\gamma}^{\dag}(-\omega)$, which results in
\begin{equation}
\hat{b}_{\gamma}^{\dag} (\omega) = \frac{\sqrt{\Gamma} \hat{b}^{\dag}_n(\omega)}{-i \left( \omega_0 + \omega \right) + \Gamma / 2}.
\label{bofwdag}
\end{equation}
\noindent Using these results for $\hat{b}_\gamma(\omega)$ and $\hat{b}^\dag_\gamma(\omega)$, we find a damped representation of the position operator to be
\begin{equation}
\hat{x}_\gamma(\omega) = x_{\rm zpf} \left( \hat{b}_\gamma(\omega) + \hat{b}^\dag_\gamma(\omega) \right).
\label{xofw}
\end{equation}

Finally, with our definition of the inverse Fourier transform, along with Eq.~(\ref{noisecorrt}), we obtain the correlators for the Fourier transforms of the bath operators in frequency space as
\begin{equation}
\begin{split}
\braket{\hat{b}_n(\omega) \hat{b}^{\dag}_n(\omega') } &= 2 \pi \left( n_b + 1 \right) \delta(\omega + \omega'), \\
\braket{\hat{b}^{\dag}_n(\omega) \hat{b}_n(\omega') } &= 2 \pi n_b  \delta(\omega+\omega'), \\
\braket{\hat{b}^{\dag}_n(\omega) \hat{b}^{\dag}_n(\omega') } &=  \braket{\hat{b}_n(\omega) \hat{b}_n(\omega') } = 0 .
\label{noisecorrw}
\end{split}
\end{equation}
\noindent Note that a difference of a factor of $2 \pi$ arises between these correlators and others found in the literature \cite{gardiner,safavi-naeini} due to our definition of the Fourier transform. These operators with damping included will be useful later when determining the PSD for the damped quantum harmonic oscillator.

\section{Autocorrelation Functions and Power Spectral Densities}
\label{ACFPSD}

In this section, we provide definitions that allow us to calculate the ACFs and PSDs for classical signals and quantum operators. Note that in this document, we will introduce a bar over the classical ACFs and PSDs to differentiate them from their quantum analogs.

\subsection{Classical}
\label{ACFPSDcl}

We begin with a classical description of the ACF for a real, time-dependent signal $a(t)$. The ACF tells us how the value of $a(t)$ at a time $t'$ is correlated to itself at a later time $t + t'$ and is given by \cite{norton}
\begin{equation}
\bar{R}_{aa}(t) = \displaystyle  \lim_{T_0 \to \infty} \frac{1}{T_0} \int_{-\infty}^{\infty} \! a(t') a(t' + t) \,  dt'.
\label{ACFdefcl}
\end{equation}
Furthermore, by taking $t=0$, that is inspecting how $a(t)$ is related to itself at the same time, we obtain the time average of $a^2(t)$ defined as
\begin{equation}
\left< a^2 \right> = \bar{R}_{aa}(0) = \displaystyle  \lim_{T_0 \to \infty} \frac{1}{T_0} \int_{-\infty}^{\infty} \! a^2(t') \, dt' ,
\label{a2defACFcl}
\end{equation}
\noindent where we have used the shorthand $\left< a^2 \right> = \left< a^2(t) \right>$ and will continue to use this notation throughout the document. 

The PSD, which specifies the signal's intensity at a given frequency, and the ACF for a signal are related to each other by a Fourier transform. Therefore we can obtain the PSD for $a(t)$ from its ACF by \cite{norton}
\begin{equation}
\bar{S}_{aa}(\omega) = \int_{-\infty}^{\infty} \! \bar{R}_{aa}(t) e^{i \omega t} \,  dt. 
\label{PSDdefcl}
\end{equation}
\noindent Furthermore, we can use the properties of the Fourier transform given in \ref{four}, along with the definition of the ACF from Eq.~(\ref{ACFdefcl}) to write this PSD in terms of the Fourier transform of $a(t)$ as
\begin{equation}
\bar{S}_{aa}(\omega) = \displaystyle \lim_{T_0 \to \infty} \frac{1}{T_0} \left| a(\omega) \right|^2,
\label{PSDfoura}
\end{equation}
\noindent where $a(\omega) = \mathcal{F} \{ a(t) \}$. By performing the inverse Fourier transform we can also recover the ACF from the PSD as
\begin{equation}
\bar{R}_{aa}(t) = \frac{1}{2 \pi} \int_{-\infty}^{\infty} \! \bar{S}_{aa}(\omega) e^{-i \omega t} \,  d \omega.
\label{ACFinvfourcl}
\end{equation}
\noindent Also, through Eq.~(\ref{a2defACFcl}) it is apparent that the PSD is related to the time average of the squared signal by
\begin{equation}
\left< a^2 \right> = \frac{1}{2 \pi} \int_{-\infty}^{\infty} \! \bar{S}_{aa}(\omega) \, d \omega .
\label{a2defPSDcl}
\end{equation}
\noindent Generally, the energy of the signal is proportional to the signal itself squared, so by integrating the PSD over all frequencies, we are able to determine the average energy of the signal in question. This property will be useful later when normalizing our PSDs. 

We conclude our discussion on the classical PSD and ACF by noting that the definitions we have chosen are for the two-sided PSD, which is defined for both positive and negative frequencies. We elect to use the classical two-sided PSD for this document, as it is easier to correspond with the quantum PSD, in which an asymmetry between positive and negative frequency arises.  However, we mention briefly that when performing classical experiments, it is sometimes more convenient to work with the one-sided displacement PSD, which is defined over only positive frequencies \cite{press} and is often quoted in the literature \cite{hauer,albrecht,ekinci,krause}. Using the fact that a classical two-sided PSD is an even function, we can see that in performing the integrals in Eqs.~(\ref{ACFinvfourcl}) and (\ref{a2defPSDcl}), the limits can be changed from 0 to $\infty$, provided we multiply by a factor of 2. Therefore, we can determine the one-sided PSD by multiplying the two-sided PSD by a factor of two and restricting its definition to be over only positive frequencies. This simple conversion from a two-sided to a one-sided PSD applies to all classical PSDs derived in this document.

\subsection{Quantum}
\label{ACFPSDqu}

In the realm of quantum mechanics, physical observables correspond to Hermitian operators that act on wavefunctions. Therefore, our ACF and PSD will be in terms of the averages of these operators. 

The quantum PSD is a spectral function that tells us the intensity of a time-dependent quantum mechanical operator $\hat{a}(t)$ at a given frequency $\omega$ and is defined as \cite{clerk2}
\begin{equation}
\begin{split}
S_{aa}(\omega) &= \int^{\infty}_{-\infty} \! R_{aa}(t)  e^{i \omega t} \, dt \\ 
&= \int^{\infty}_{-\infty} \! \left< \hat{a}(t) \hat{a}(0) \right>  e^{i \omega t} \, dt,
\label{PSDdefqu}
\end{split}
\end{equation}
\noindent where $R_{aa}(t) = \left< \hat{a}(t) \hat{a}(0) \right>$ is the ACF for $\hat{a}(t)$. At a finite temperature $T$, we can determine the ACF for $\hat{a}(t)$ from
\begin{equation}
\left< \hat{a}(t) \hat{a}(0) \right> = \frac{\text{Tr} \{ e^{-\beta \hat{H}} e^{i \hat{H} t / \hbar} \hat{a} e^{-i \hat{H} t / \hbar} \hat{a} \}}{\text{Tr} \{ e^{-\beta \hat{H}} \}},
\label{ACFdefqu}
\end{equation}
\noindent where $\hat{H}$ is the Hamiltonian of the system, $\beta = 1/k_B T$ with $k_B$ being the Boltzmann constant and Tr$\{ \}$ denotes the trace of an operator. In this paper, we choose to work in the energy eigenstate basis so that the trace of an operator $\hat{O}$ is given by
\begin{equation}
\text{Tr} \{ \hat{O} \} = \displaystyle\sum\limits_{n} \bra{n} \hat{O} \ket{n},
\label{tracedef}
\end{equation}
\noindent where $\ket{n}$ is the $n$th energy eigenstate of our quantum system. We can therefore see that the denominator of Eq.~(\ref{ACFdefqu}), given by
\begin{equation}
Z = \displaystyle\sum\limits_{n} \bra{n} e^{-\beta \hat{H}} \ket{n} = \displaystyle\sum\limits_{n} e^{-\beta E_n} \braket{n|n} = \displaystyle\sum\limits_{n} e^{-\beta E_n},
\label{partfuncdef}
\end{equation}
\noindent is the canonical partition function \cite{gardiner}.

We also point out that we can inverse Fourier transform $S_{aa}(\omega)$ to obtain $R_{aa}(t)$ as
\begin{equation}
R_{aa}(t) = \left< \hat{a}(t) \hat{a}(0) \right> = \frac{1}{2 \pi} \int_{-\infty}^{\infty} \! S_{aa}(\omega) e^{-i \omega t} \, d \omega.
\label{ACFinvfourqu}
\end{equation}
\noindent Setting $t=0$ we then have
\begin{equation}
\braket{\hat{a}^2} = \frac{1}{2 \pi} \int_{-\infty}^{\infty} \! S_{aa}(\omega) \, d \omega,
\label{a2defPSDqu}
\end{equation}
\noindent in direct correspondence with Eq.~(\ref{a2defPSDcl}) for a classical signal.

It is also possible to express $S_{aa}(\omega)$ in terms of the Fourier transform of $\hat{a}(t)$. By inputting the definitions for the Fourier transform of $\hat{a}(t)$ into Eq.~(\ref{PSDdefqu}) we find
\begin{equation}
S_{aa}(\omega) = \frac{1}{2 \pi} \int^{\infty}_{-\infty} \! \left< \hat{a}(\omega) \hat{a}(\omega') \right> \,  d\omega',
\label{PSDdefwqu}
\end{equation}
\noindent where $\hat{a}(\omega) = \mathcal{F} \{ \hat{a}(t) \}$. This relation is very useful, as it provides an alternate method by which we can calculate PSDs using the frequency domain. Equipped with these definitions, as well as the relations given in Section \ref{back}, we are now ready to determine the PSDs for the harmonic oscillator.

\section{General Formulation for the Power Spectral Density of ${\bf x^{k}}$}
\label{PSDk}

We now introduce a method by which the PSD can be calculated in both the classical and quantum regimes for any power of the position of a harmonic oscillator $x^k(t)$, where $k$ is any positive integer. From this point forth, we label the PSD for $x^k(t)$ as the $k$th-order PSD and likewise for the corresponding ACF.

\subsection{Classical}
\label{PSDkcl}

Beginning with the $k$th-order classical PSD, we use the definition of the PSD as the Fourier transform of the ACF given in Eq.~(\ref{PSDdefcl}), along with Eq.~(\ref{PSDfoura}), to obtain
\begin{equation}
\bar{S}_{x^k x^k}(\omega) = \mathcal{F} \{ \bar{R}_{x^k x^k}(t) \} = \displaystyle  \lim_{T_0 \to \infty} \frac{1}{T_0} | x^{(k)}(\omega)|^2.
\label{PSDkdefcl}
\end{equation}
\noindent Here we have used the notation $x^{(k)}(\omega) = \mathcal{F} \{x^k(t) \}$ to denote the Fourier transform of $x^k(t)$. Using Eq.~(\ref{fourprodk}), we can express this quantity as
\begin{equation}
\begin{split}
x^{(k)}(\omega) &= \mathcal{F} \{ x(t) \cdot x(t) \cdot ... \cdot x(t) \} \\
 &= \frac{1}{(2 \pi )^{k-1}} x(\omega) * x(\omega) * ... * x(\omega),
\label{xkwcl}
\end{split}
\end{equation}
\noindent where the ellipsis (...) indicates that the corresponding operation is performed on $k$ terms (for a total of $k-1$ operations). We point out that with this notation, $k=1$ corresponds to a single term with no operations performed. 

For a general driving force $f(\omega)$, this expression is very difficult to solve. However, if we restrict ourselves to a frequency-independent drive ({\it i.e.}~$f(\omega) = F$), as is the case in thermally driven classical oscillators, the problem simplifies significantly, as we obtain the relation
\begin{equation}
 x(\omega) * x(\omega) * ... * x(\omega) = F^k \chi(\omega) * \chi(\omega) * ... * \chi(\omega),
\label{xkchi}
\end{equation}
\noindent where we have input the relation in Eq.~(\ref{CDHOfreq}) for $x(\omega)$. We can then write
\begin{equation}
\bar{S}_{x^k x^k}(\omega) = \bar{S}^{\rm th}_{F^k F^k} \left| \chi(\omega) * \chi(\omega) * ... * \chi(\omega) \right|^2,
\label{PSDkchi}
\end{equation}
\noindent where we have defined a white noise thermal force PSD
\begin{equation}
\bar{S}^{\rm th}_{F^k F^k} = \displaystyle \lim_{T_0 \to \infty} \frac{1}{T_0} \left| \frac{F^k}{(2 \pi)^{k-1}} \right|^2.
\label{PSDFkFk}
\end{equation}
\noindent Note that while the driving force is constant in frequency space, it still grows as we increase $T_0$, balancing out the division by infinity such that $\bar{S}^{\rm th}_{F^k F^k}$ remains constant. The value of this quantity can be determined by ensuring that Eq.~(\ref{a2defPSDcl}) is satisfied. In the high-$Q$ limit, we can approximate the expectation value of $\left< x^{2k} \right>$ for a damped harmonic oscillator as that for the undamped oscillator in equilibrium with a bath at temperature $T$ (see \ref{x2kave}), which results in
\begin{equation}
\left< x^{2k} \right> = \frac{1}{2 \pi} \int_{-\infty}^{\infty} \! \bar{S}_{x^k x^k}(\omega) \, d \omega = x_{\rm th}^{2k}\frac{(2k)!}{2^k k!},
\label{x2kavecl}
\end{equation}
\noindent where we have introduced the root-mean-square amplitude of our thermally driven motion as $x_{\rm th} = \sqrt{1/\beta m \omega_0^2} = \sqrt{k_B T / m \omega_0^2}$. In order to satisfy this normalization condition, we must integrate over the PSD once we have determined its functional form by evaluating the convolutions found in Eq.~(\ref{PSDkchi}).

Before we move on to the quantum PSD, we provide a brief remark in regards to carrying out the above procedure. As can be seen above, the calculations performed using this method become increasingly tedious as $n$ becomes larger, mainly due to the increasing number of convolutions. However, this complexity can be alleviated slightly by breaking up the convolutions into smaller calculations, allowing us to calculate our PSDs in an iterative manner which utilizes previous calculations. For instance, if we have already determined the second-order PSD, for which we need $\chi(\omega) * \chi(\omega)$, we can convolve this quantity with $\chi(\omega)$, or with itself, and use the result to determine the third- and fourth-order PSDs, respectively, reducing the number of convolutions needed.

\subsection{Quantum}
\label{PSDkqu}

We now move onto calculation of the quantum PSD for $\hat{x}^k(t)$. In this case, it is easier to focus on calculating the ACF, which can then be Fourier transformed to produce the corresponding PSD. Using Wick's Theorem \cite{mahan,fetter}, we are able to determine the $k$th-order ACF to be (see \ref{Wick})
\begin{equation}
\begin{split} 
R_{x^k x^k}(t) &= \braket{\hat{x}^k(t) \hat{x}^k(0)} \\
&= \displaystyle\sum\limits_{c=0}^{N} A_c\braket{\hat{x}(t) \hat{x}(t)}^c \braket{\hat{x}(t)\hat{x}(0)}^{k - 2c} \braket{\hat{x}(0) \hat{x}(0)}^c,
\label{ACFkdefqu1}
\end{split}
\end{equation}
\noindent where 
\begin{equation} 
A_c= \frac{\left( k! \right)^2}{2^{2c} \left(c! \right)^2 \left( k - 2c \right)!},
\label{ACFkAc}
\end{equation}
\noindent and
\begin{equation} 
N = \begin{cases} k/2 &\mbox{for even } k, \\ 
(k-1)/2 & \mbox{for odd } k. \end{cases} 
\label{ACFkN}
\end{equation}
\noindent Therefore, by using Wick's theorem, we have reduced the complex problem of finding a $2k$ term correlation function to evaluating the two term correlation functions under the sum in Eq.~(\ref{ACFkdefqu1}), which we determine to be (see \ref{2pACF})
\begin{equation} 
\begin{split}
\braket{\hat{x}(t)\hat{x}(0)} &=  x_{\rm zpf}^2\left[ \left( \braket{n} + 1 \right)e^{-i \omega_0 t} + \braket{n} e^{i \omega_0 t} \right], \\
\braket{\hat{x}(t) \hat{x}(t)} &= \braket{\hat{x}(0) \hat{x}(0)} = x_{\rm zpf}^2 \left[ 2 \braket{n} + 1 \right]. 
\label{ACFk2pcorr}
\end{split}
\end{equation}
\noindent Here we point out that in the first line we have obtained the expression for the ACF in the $k=1$ (linear) case. In the above equations, we have introduced the thermal average of $n$ for the harmonic oscillator, which is given by
\begin{equation}
\braket{n} = \frac{1}{e^{\beta \hbar \omega_0} - 1}.
\label{aven}
\end{equation}
\noindent This quantity can be interpreted as the average number of quanta obeying Bose-Einstein statistics at a temperature determined by $\beta$. Combining the results of Eq.~(\ref{ACFk2pcorr}) with Eq.~(\ref{ACFkdefqu1}) we obtain
\begin{equation}
\begin{split}
R_{x^k x^k}(t) &= x_{\rm zpf}^{2k} \displaystyle\sum\limits_{c=0}^{N} A_c \left( 2 \braket{n} + 1 \right)^{2c} \\ 
&\times \left[ \left( \braket{n} + 1 \right)e^{-i \omega_0 t} + \braket{n} e^{i \omega_0 t} \right]^{k - 2c}.
\label{ACFkdefqu2}
\end{split}
\end{equation}
\noindent Using the binomial theorem, we can instead write our $k$th-order ACF in the form
\begin{equation}
R_{x^k x^k}(t) = x_{\rm zpf}^{2k} \displaystyle\sum\limits_{c=0}^{N} \displaystyle\sum\limits_{d=0}^{k-2c} B_{cd} e^{i(2c + 2d-k)\omega_0 t},
\label{ACFkdefqu3}
\end{equation}
\noindent where we have the new coefficient
\begin{equation}
B_{cd} = \frac{\left(k!\right)^2 \left( 2 \braket{n} + 1 \right)^{2c} \left( \braket{n} + 1 \right)^{k-2c-d} \braket{n}^d}{2^{2c} (c!)^2 d! \left( k - 2c - d \right)!}.
\label{ACFkBcd}
\end{equation}
\noindent In this form, we can easily Fourier transform Eq.~(\ref{ACFkdefqu3}) to obtain the $k$th-order quantum PSD
\begin{equation}
\begin{split}
S_{x^k x^k}(\omega) &= \int_{-\infty}^{\infty} \! R_{x^k x^k}(t) e^{i \omega t} \, dt \\ 
&= 2 \pi x_{\rm zpf}^{2k} \displaystyle\sum\limits_{c=0}^{N} \displaystyle\sum\limits_{d=0}^{k-2c} B_{cd} \delta\left( \omega + (2c + 2d-k)\omega_0 \right),
\label{PSDkdefqu}
\end{split}
\end{equation}
\noindent where we have used the definition of the Dirac delta function given by Eq.~(\ref{deltafour}). This provides an expression for the $k$th-order PSD for the undamped quantum harmonic oscillator for any positive integer $k$. 

In any realistic system, however, a non-zero amount of damping will occur as the oscillator radiates energy to its environment. To determine the the $k$th-order PSD with damping included, we could in principle use Eq.~(\ref{PSDdefwqu}) to calculate our PSD according to
\begin{equation}
\tilde{S}_{x^k x^k}(\omega) = \frac{1}{2 \pi} \int_{-\infty}^{\infty} \! \braket{ \hat{x}_\gamma^{(k)}(\omega) \hat{x}_\gamma^{(k)}(\omega')} \, d \omega',
\label{PSDkdampdef}
\end{equation}
\noindent where $\hat{x}^{(k)}(\omega) = \mathcal{F} \{ \hat{x}^k(t) \} = \hat{x}(\omega)*...*\hat{x}(\omega)/ (2 \pi)^{k-1}$. Here we have included a tilde over this PSD symbol to indicate that it is a quantum PSD with damping included. As we can see here, because we must work in the frequency domain for the input-output formalism of the damped harmonic oscillator, to determine the quantities $\hat{x}^{(k)}(\omega)$ we must compute $k-1$ convolution integrals. Just like in the classical case, this leads to an increasingly complex problem as we increase $k$.

Fortunately, by using a definition of the delta function, we have an alternate method by which we can include damping into the $k$th-order PSD. In the case of small $\Gamma$, we can approximate the delta functions in Eq.~(\ref{PSDkdefqu}) using Eq.~(\ref{deltalor}) to obtain
\begin{equation}
\delta_k(\omega) \approx \frac{1}{2 \pi} \frac{k\Gamma}{\omega^2 + \left(k\Gamma/2\right)^2}.
\label{deltak}
\end{equation}
\noindent The subscript $k$ is added here to differentiate between delta functions of different orders, as the half-width of the peaks of the PSD increases as $k \Gamma / 2$. This effect is discussed in detail in Section \ref{PSD2cl}. Using the expression in Eq.~(\ref{deltak}), we can write our $k$th-order damped PSD as
\begin{equation}
\tilde{S}_{x^k x^k}(\omega) = x_{\rm zpf}^{2k} \displaystyle\sum\limits_{c=0}^{N} \displaystyle\sum\limits_{d=0}^{k-2c} \frac{ k \Gamma B_{cd}}{\left( \omega + (2c + 2d-k)\omega_0 \right)^2 + \left( k \Gamma/2 \right)^2},
\label{PSDkdampqu}
\end{equation}
\noindent where we have a sum of Lorentzians instead of delta functions, effectively introducing damping into our quantum PSD. We will show below that for the $k=1$ case, this result is exactly what would be obtained if we had instead decided to use the input-output formalism to include damping in our system, justifying this simpler approach.

In concluding this section, we would like to point out that it is possible to use our result for the $k$th-order ACF to determine the thermal average of $\hat{x}^{2k}(t)$. Taking $t=0$ in Eq.~(\ref{ACFkdefqu1}) we have
\begin{equation}
\braket{\hat{x}^{2k}} = x_{\rm zpf}^{2k} \left( 2 \braket{n} + 1 \right)^k \frac{(2k)!}{2^k k!} = \left( \frac{\braket{\hat{H}}}{m \omega_0^2} \right)^k \frac{(2k)!}{2^k k!},
\label{x2kavequ}
\end{equation}
\noindent where $\braket{\hat{H}} = \hbar \omega_0 ( \braket{n} + 1/2 )$ is the average energy of the harmonic oscillator. This equation, unlike the classical analog, is valid for all temperatures as demonstrated by the fact that Eq.~(\ref{x2kavecl}) is recovered by taking the high temperature limit $k_B T \gg \hbar \omega_0$, for which $2 \braket{n} + 1 \approx 2 k_B T / \hbar \omega_0$. 

The $T = 0$ limit of Eq.~(\ref{x2kavequ}) can also be taken. Upon inspection of Eq.~(\ref{aven}), we see that as $T \rightarrow 0$, $\braket{n} \rightarrow 0$ indicating that the oscillator is in its ground state, giving
\begin{equation}
\braket{\hat{x}^{2k}} =  x_{\rm zpf}^{2k} \frac{(2k)!}{2^k k!}.
\label{x2kaveT0}
\end{equation}
\noindent We point out that this equation provides a quantum analog to Eq.~(\ref{x2kavecl}), where we have taken $x_{\rm th} \rightarrow x_{\rm zpf}$, as our system is purely driven by quantum fluctuations in the ground state as opposed to the classical thermal drive.

\section{First-Order Power Spectral Density}
\label{PSD1}

Now that the framework for determining PSDs and ACFs for the harmonic oscillator has been laid out, we show that for $k=1$ our formalism reproduces the well-known results of the first-order PSD for the position of the harmonic oscillator in both the classical and quantum regimes.

\subsection{Classical}
\label{PSD1cl}

The linear displacement PSD for the classical damped harmonic oscillator is determined by taking $k=1$ in Eq.~(\ref{PSDkchi}), producing
\begin{equation}
\begin{split}
\bar{S}_{xx}(\omega) = \bar{S}^{\rm th}_{FF}(\omega)  \left| \chi(\omega) \right|^2 = \frac{\bar{S}_{FF}^{\rm th}}{m^2 \left[ (\omega^2 - \omega_0^2)^2 + \omega^2 \Gamma^2 \right]},
\end{split}
\label{PSD1defcl}
\end{equation}
\noindent where we have used the generalized mechanical susceptibility found in Eq.~(\ref{mechchi}). In order to determine the constant $\bar{S}_{FF}^{\rm th}$, we integrate $\bar{S}_{xx}(\omega)$ over all frequencies (see \ref{PSD1contint}) and use Eq.~(\ref{a2defPSDcl}) to obtain
\begin{equation}
\left< x^2 \right> = \frac{\bar{S}_{FF}^{\rm th}}{2 \pi m^2} \int_{-\infty}^{\infty} \! \frac{d \omega}{(\omega^2 - \omega_0^2)^2 + \omega^2 \Gamma^2} =  \frac{\bar{S}_{FF}^{\rm th}}{2 m^2 \omega_0^2 \Gamma }.
\label{x2avedefcl}
\end{equation}
\noindent Inputting $k=1$ into Eq.~(\ref{x2kavecl}) we also have
\begin{equation}
\left< x^2 \right> = \frac{k_B T}{m \omega_0^2}.
\label{x2averescl}
\end{equation}
\noindent This result, which can be written in the form 
\begin{equation}
\frac{1}{2} m \omega_0^2 \braket{x^2} = \braket{V} = \frac{1}{2} k_B T,
\label{equippartcl}
\end{equation}
\noindent is simply the equipartition theorem for the classical harmonic oscillator in thermal equilibrium at a temperature $T$, for which the average potential energy $V$ is equal to $k_B T / 2$ \cite{bowley}.

By equating Eqs.~(\ref{x2avedefcl}) and (\ref{x2averescl}), we find $\bar{S}_{FF}^{\rm th} = 2 m \Gamma k_B T$, which allows us to write the displacement PSD for the classical damped harmonic oscillator as
\begin{equation}
\bar{S}_{xx}(\omega) = \frac{2 \Gamma k_B T}{m \left[ (\omega^2 - \omega_0^2)^2 + \omega^2 \Gamma^2 \right]}.
\label{PSD1fincl}
\end{equation}
\noindent This result agrees with that found in the literature \cite{hauer,albrecht,ekinci,krause}, provided we incorporate the factor of 2 required when transferring between one- and two-sided PSDs. The above result could have also been obtained in a more straightforward route using the classical fluctuation-dissipation theorem \cite{landau}, which states that
\begin{equation}
\bar{S}_{xx}(\omega) = \frac{2 k_B T}{\omega} {\rm Im} \{ \chi(\omega) \}.
\label{PSD1FDTcl}
\end{equation}

With the functional form of our PSD, we are now able to investigate some of its properties. First, since the signal is peaked at the resonance frequency and the PSD is an even function, we know that peaks exist at $\omega = \pm \omega_0$, which results in
\begin{equation}
\bar{S}_{xx}^{\rm max} = \bar{S}_{xx}(\pm \omega_0) = \frac{2 k_B T}{m \omega_0^2 \Gamma}.
\label{PSD1max}
\end{equation}

Another interesting parameter of the PSD is the width of the peak, which is closely related to the damping of the oscillator. Here, we consider the full width at half maximum (FWHM) $\Delta \omega$. To determine this quantity we look for the frequencies $\omega_{1/2}$ at which $\bar{S}_{xx}(\omega_{1/2}) = \bar{S}_{xx}^{\rm max} /2$, which leads to the quartic equation
\begin{equation}
\omega_{1/2}^4 + \left( \Gamma^2 - 2 \omega_0^2 \right) \omega_{1/2}^2 + \omega_0^2 \left( \omega_0^2 - 2 \Gamma^2 \right) = 0.
\label{PSD1halffreqeq}
\end{equation}
\noindent Using the quadratic formula, the solutions to this equation are found to be
\begin{equation}
\omega_{1/2} = \pm \sqrt{ \omega_0^2 - \frac{\Gamma^2}{2} \pm \frac{\Gamma^2}{2} \sqrt{ \Gamma^2 + 4 \omega_0^2}} \approx \pm \omega_0 \pm \frac{\Gamma}{2},
\label{PSD1halffreq}
\end{equation}
\noindent where we have made the high-$Q$ approximation. The four solutions in the above equation correspond to two points on the sides of the two peaks at $\pm \omega_0$, which leads to a FWHM of $\Delta \omega = \Gamma$ in the high-$Q$ limit.

\subsection{Quantum}
\label{PSD1qu}

Moving to the quantum regime, we now look to determine the first-order ACF and PSD for the position operator of the quantum harmonic oscillator. We have already calculated the first-order ACF in Section \ref{PSDkqu} where it was found to be (see the first line of Eq.~(\ref{ACFk2pcorr}))
\begin{equation}
\begin{split}
R_{xx}(t) &= \left< \hat{x}(t) \hat{x}(0) \right> \\
&= x_{\rm zpf}^2 \left[ \left( \langle n \rangle +1 \right) e^{-i \omega_0 t} + \langle n \rangle e^{i \omega_0 t} \right].
\label{ACF1defqu}
\end{split}
\end{equation}
\noindent By Fourier transforming this ACF, or equivalently taking $k=1$ in Eq.~(\ref{PSDkdefqu}), we obtain
\begin{equation}
S_{xx}(\omega) = 2 \pi  x_{\rm zpf}^2 \left[  (\braket{n} + 1) \delta (\omega - \omega_0) + \braket{n} \delta (\omega + \omega_0) \right].
\label{PSD1finqu}
\end{equation}
\noindent This produces the well-known expression for the first-order PSD for the position operator of the quantum harmonic oscillator \cite{clerk2}. This result is also verified by an independent determination using the fluctuation-dissipation theorem (see \ref{PSDFDT}).

\begin{figure}[h!]
\includegraphics[width = \columnwidth]{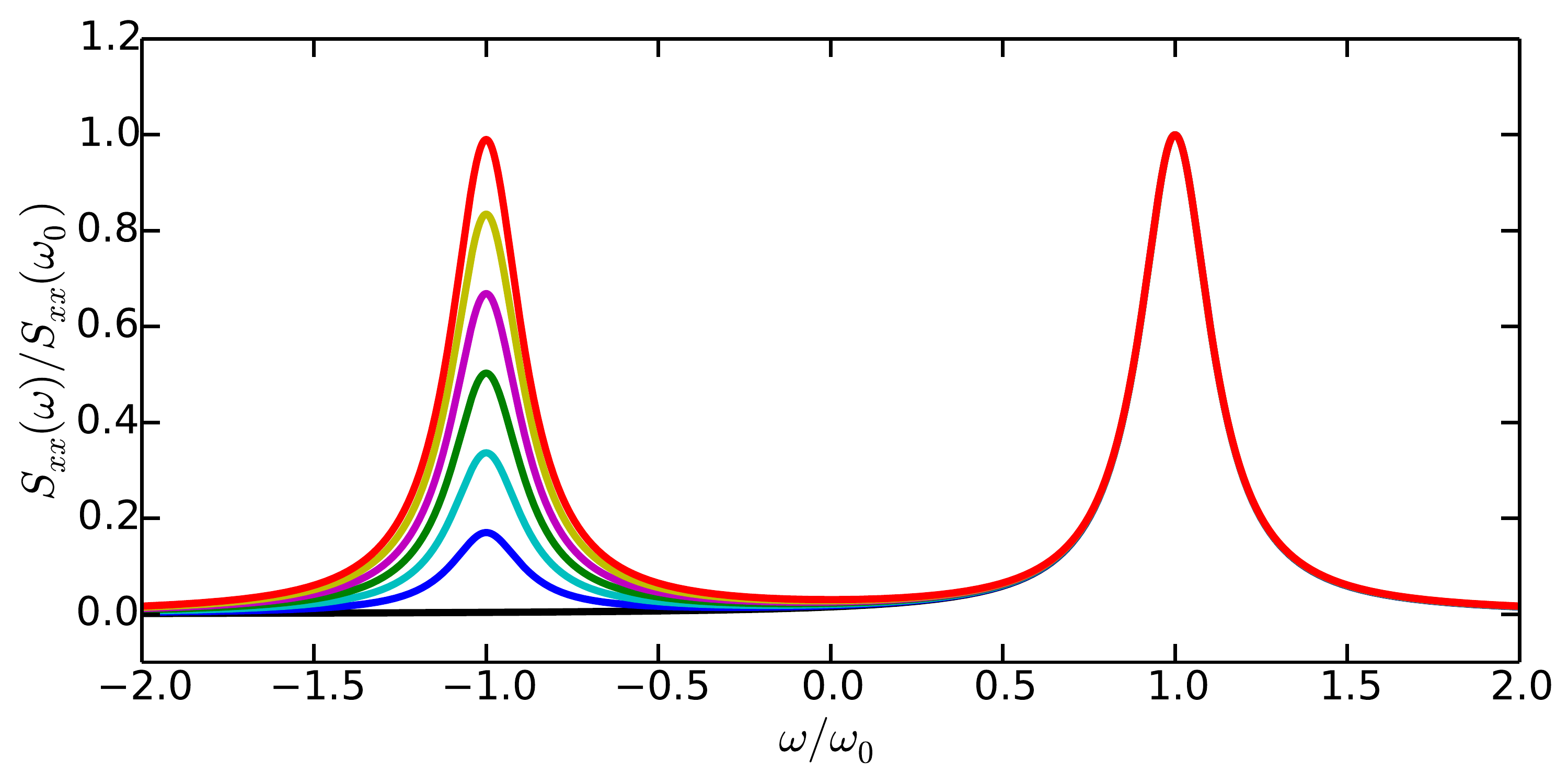}
\caption{The first-order damped quantum PSD, with each trace normalized such that its maximum at $\omega = \omega_0$ is 1, vs frequency in terms of the resonant frequency. We have chosen a relatively low quality factor of 4 for this PSD, so that the amplitude effects are not obscured by the narrowness of the peaks. The different colors represent different average quanta with $\braket{n}$ = 0, 0.2, 0.5, 1, 2, 5, 100, as we move up in color from black to red. While the relative height of the peak at $\omega=\omega_0$ is unchanged, the peak at $\omega=-\omega_0$ decreases with average number of quanta, demonstrating the asymmetry of the quantum PSD. In the extreme case of $\braket{n} = 0$, the $\omega = -\omega_0$ peak disappears altogether. }
\label{1stOrderAsymm}
\end{figure}

The first-order quantum PSD is not symmetric about zero frequency as was the case for the first-order classical PSD. This asymmetry is visualized in Fig.~\ref{1stOrderAsymm}. Physically, these two peaks correspond to two different processes. The negative frequency peak ($\omega = -\omega_0$) is associated with the annihilation/emission of a single quantum with frequency $\omega_0$. Alternately, the positive frequency peak corresponds to the creation/absorption of a quantum at $\omega_0$. In the context of optomechanics, these processes are strongly tied to Stokes/anti-Stokes Raman scattering whereby phonons can be created/annihilated via interaction with cavity photons \cite{aspelmeyer}. Furthermore, the asymmetry of these peaks leads to distinctly non-classical effects at low phonon number, such as motional sideband asymmetry, which has recently been observed experimentally \cite{safavi-naeini2,weinstein}. 

It is also interesting to investigate the $T = 0$ limit of the above quantum PSD. This limit corresponds to the PSD of a quantum harmonic oscillator that is purely in its ground state, its motion arising solely from zero point fluctuations due to quantum noise. Taking $\braket{n} = 0$ in Eq.~(\ref{PSD1finqu}), the quantum PSD becomes
\begin{equation}
S^0_{xx}(\omega) = 2 \pi  x_{\rm zpf}^2 \delta (\omega - \omega_0).
\label{PSD1T0}
\end{equation}
\noindent In this limit, we completely lose the peak at $\omega = -\omega_0$ due to the fact that in the ground state no quanta exist to annihilate.

The above discussion on the physical significance of the quantum PSD was for the ideal case of zero damping, leading to perfectly narrow peaks corresponding to quanta at two distinct resonance frequencies, $\omega = \pm \omega_0$. In a realistic system, however, damping will emerge, broadening these peaks and allowing for small deviations from this resonance frequency. We now look to include damping into our system by using the input-output formalism outlined in Section \ref{QHO}. In this case, we calculate the PSD using Eq.~(\ref{PSDdefwqu}) to obtain
\begin{equation}
\tilde{S}_{xx}(\omega) = \frac{1}{2 \pi} \int_{-\infty}^{\infty} \! \braket {\hat{x}_\gamma(\omega) \hat{x}_\gamma(\omega') } \, d \omega' .
\label{PSD1dampdefqu}
\end{equation}
\noindent Utilizing Eq.~(\ref{xofw}), we find this damped PSD to be (see \ref{PSDdamp})
\begin{equation}
\tilde{S}_{xx}(\omega) = \Gamma x_{\rm zpf}^2 \Bigg[ \frac{\braket{n}+1}{\left(\omega - \omega_0 \right)^2 + \left( \Gamma / 2 \right)^2} + \frac{\braket{n}}{\left(\omega + \omega_0 \right)^2 + \left( \Gamma / 2 \right)^2} \Bigg],
\label{PSD1dampfinqu}
\end{equation}
\noindent which also agrees to what is found in the literature \cite{clerk2,safavi-naeini2}. A $T=0$ PSD corresponding to the ground state of a damped harmonic oscillator can also be determined by setting $\braket{n} = 0$ in the above equation to obtain
\begin{equation}
\tilde{S}^0_{xx}(\omega) = \frac{ \Gamma x_{\rm zpf}^2}{\left(\omega - \omega_0 \right)^2 + \left( \Gamma / 2 \right)^2}.
\label{PSD1dampT0}
\end{equation}
\noindent Both of these above results could have also been obtained by simply taking $k=1$ in Eq.~(\ref{PSDkdampqu}), justifying the method by which we obtained this expression.

Finally, we can also find the thermal average $\left< \hat{x}^2 \right>$ for the quantum harmonic oscillator using Eq.~(\ref{a2defPSDqu}), where we can integrate over either $S_{xx}(\omega)$ or $\tilde{S}_{xx}(\omega)$ to obtain
\begin{equation}
\left< \hat{x}^2 \right> =  x_{\rm zpf}^2(2 \braket{n} + 1)  = \frac{\braket{\hat{H}}}{m\omega_0^2},
\label{x2avequ}
\end{equation}
\noindent consistent with $k=1$ in Eq.~(\ref{x2kavequ}). In the ground state, we then have $\left< \hat{x}^2 \right> = x_{\rm zpf}^2$ such that the average value of the squared motion is the zero point fluctuation amplitude squared, as would be expected. This above equation can also be recast into
\begin{equation}
\frac{1}{2} m \omega_0^2 \left< \hat{x}^2 \right> = \frac{\hbar \omega_0}{2} \left( \braket{n} + \frac{1}{2} \right) = \frac{\braket{\hat{H}}}{2} = \braket{\hat{V}}.
\label{geneqp}
\end{equation}
\noindent This can be interpreted as a sort of ``generalized'' equipartition theorem in which the average potential energy can be related to the average value of position squared, regardless of whether the drive results from thermal or quantum noise. As such, by taking the high temperature limit of this equation, the classical equipartition partition theorem given in Eq.~(\ref{equippartcl}) is recovered.

\subsection{Classical Correspondence}
\label{PSD1cc}

For any quantum mechanical model, the correspondence principle tells us that the quantum result will reproduce its classical analog when the appropriate limits are taken. We will now show that for the first-order PSD for the damped quantum harmonic oscillator calculated above, we are able to retrieve the classical linear PSD in the limits of high temperature ($k_B T \gg \hbar \omega_0$) and quality factor ($Q \gg 1/2$). This second condition must be taken as we have implicitly made assumptions of high-$Q$ when introducing damping into the PSDs for the quantum harmonic oscillator. 

\begin{figure}[h!]
\includegraphics[width = \columnwidth]{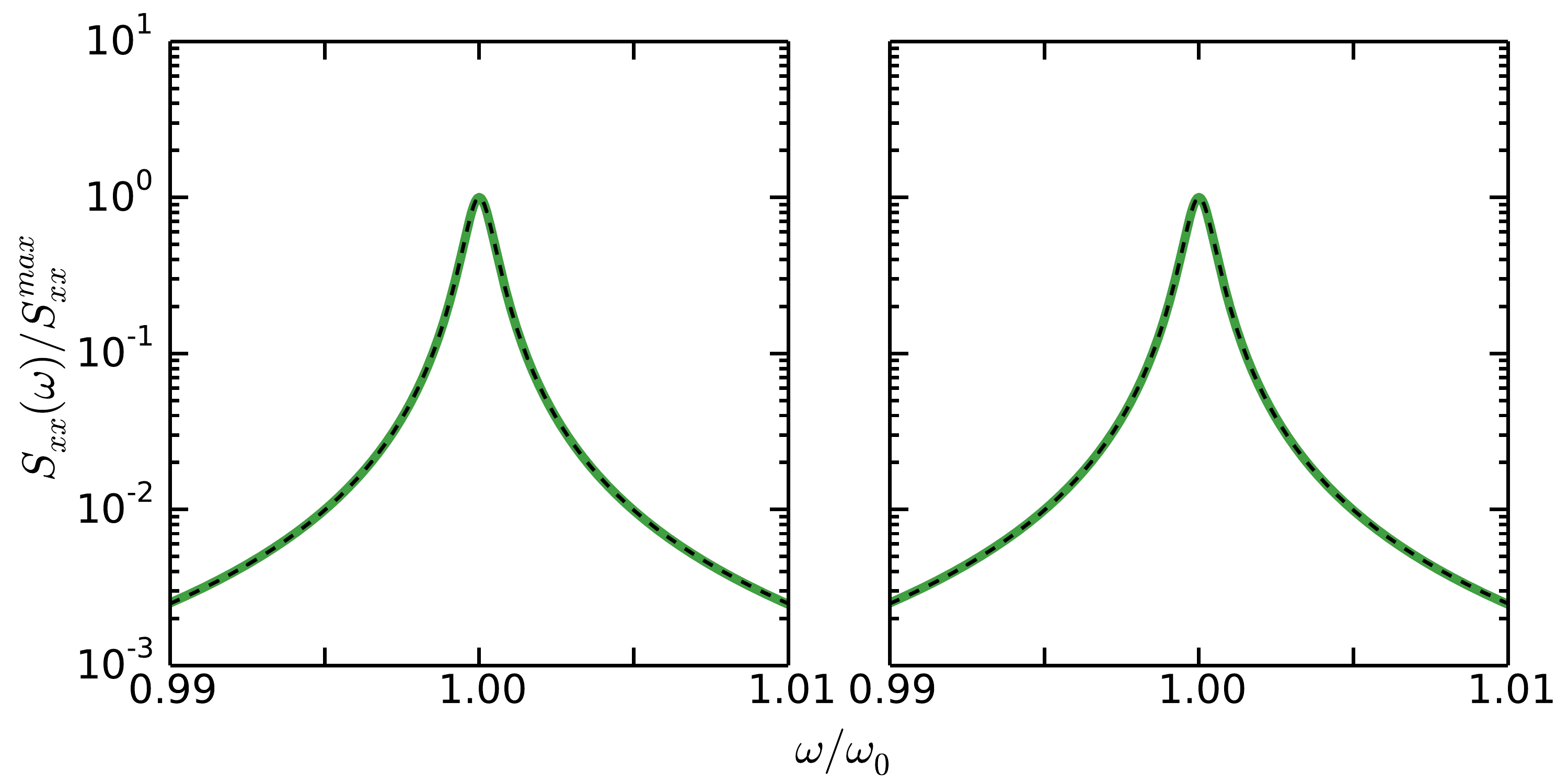}
\caption{Plot of the classical PSD (solid green) given by Eq.~(\ref{PSD1fincl}) and the damped quantum PSD (dashed black) given in Eq.~(\ref{PSD1dampfinqu}) in the regions surrounding the peaks at $\omega = \pm \omega_0$, exhibiting classical correspondence for the first order case. We have used $\braket{n} = 1000$ and $Q=1000$ such that the plots shown span $10 \Gamma$ on either side of each peak. The PSDs are normalized such that their maximum values are 1 and the frequency is given in terms of the oscillator's resonance frequency.}
\label{1stOrderQMCL}
\end{figure}

To begin, we see that in the classical limit we can use Eq.~(\ref{aven}) to make the following approximation
\begin{equation}
\braket{n}+1 \approx \braket{n} \approx \frac{1}{1 + \beta \hbar \omega_0 - 1} = \frac{1}{\beta \hbar \omega_0} = \frac{k_B T}{\hbar \omega_0}.
\label{avenapprox}
\end{equation}
\noindent Physically, this equation tells us that at high temperatures, the thermal energy of the resonator is broken into a large number of $n$ quanta, each with energy $\hbar \omega_0$, such that the ground state energy can be neglected. Remembering that $x_{\rm zpf}^2 = \frac{\hbar}{2 m \omega_0}$, we can rewrite Eq.~(\ref{PSD1dampfinqu}) as
\begin{equation}
\begin{split}
\tilde{S}_{xx}(\omega) &\approx \frac{\Gamma k_B T}{2 m \omega_0^2}\left[ \frac{1}{(\omega - \omega_0)^2 + (\Gamma / 2)^2} + \frac{1}{(\omega + \omega_0)^2 + (\Gamma / 2)^2} \right] \\
&\approx \frac{2 \Gamma k_B T}{m \left[ (\omega^2 - \omega_0^2)^2 + \omega^2 \Gamma^2 \right]}.
\label{PSD1ccapprox}
\end{split}
\end{equation}
\noindent The details of how the approximations were made to achieve the classical result are outlined in \ref{ccapp}. In the last line, we have retrieved the classical result of Eq.~(\ref{PSD1fincl}) that was determined in Section \ref{PSD1cl}. Therefore, we have shown that the first-order quantum PSD calculated here satisfies the correspondence principle in the region of interest surrounding the peaks at $\omega=\pm \omega_0$. This is illustrated in Fig.~\ref{1stOrderQMCL} for $\braket{n}=1000$ and $Q=1000$.

\section{Second-Order Power Spectral Density}
\label{PSD2}

The linear PSD calculated in the above section can be used for situations in which the displacement of an oscillator is measured directly. However, there are situations where it is useful to measure the square of the position directly \cite{thompson,nunnenkamp,gangat,huang,borkje}, in which case we need to consider the second-order PSD for the oscillator. In this section, we shall determine this quadratic PSD for both the quantum and classical cases.

\subsection{Classical}
\label{PSD2cl}

We begin by calculating the classical second-order PSD, proceeding as we did in the previous section where we now use Eq.~(\ref{PSDkchi}) with $k=2$. The PSD in this case will be given by
\begin{equation}
\bar{S}_{x^2 x^2}(\omega) = \bar{S}_{F^2 F^2}^{\rm th} \left| \chi (\omega) * \chi( \omega) \right|^2,
\label{PSD2defcl}
\end{equation}
\noindent the functional form of which is determined by the convolution
\begin{equation}
\chi (\omega) * \chi( \omega) = \int_{-\infty}^{\infty} \! \chi( \omega') \chi(\omega - \omega') \, d \omega'.
\label{chiconv}
\end{equation}
\noindent This integral can be computed using contour integration (see \ref{contintchi}) to obtain
\begin{equation}
\chi (\omega) * \chi( \omega) = \frac{-4 \pi i}{m^2  \left( \omega + i \Gamma \right) \left( \omega^2 - 4 \omega_0^2 + 2i\omega \Gamma \right)}.
\label{chiconvres}
\end{equation}
\noindent Inputting this expression into Eq.~(\ref{PSD2defcl}), we determine the unnormalized second-order PSD as
\begin{equation}
\bar{S}_{x^2 x^2}(\omega) = \frac{16 \pi^2 \bar{S}_{F^2 F^2}^{\rm th}}{m^4 \left( \omega^2 + \Gamma^2 \right) \left( \left( \omega^2 - 4 \omega_0^2 \right)^2 + 4 \omega^2 \Gamma^2 \right)}.
\label{PSD2unnormcl}
\end{equation}
\noindent Upon investigation of this function, we see that it is peaked at $\omega = \pm 2 \omega_0$, as well as $\omega =0$. This is what we expect for the PSD of the squared displacement \cite{nunnenkamp,doolin}, the physical meaning of which will become more apparent when we look at the quantum case in Section \ref{PSD2qu}.

In order to properly normalize this second-order PSD, we must determine the value of $\bar{S}_{F^2 F^2}^{\rm th}$ such that
\begin{equation}
\left< x^4 \right> = \frac{1}{2 \pi} \int_{-\infty}^{\infty} \! \bar{S}_{x^2 x^2}(\omega) \, d \omega = 3 \left( \frac{k_B T}{m \omega_0^2} \right)^2,
\label{x4avedefcl}
\end{equation}
\noindent where we have simply taken $k=2$ in Eq.~(\ref{x2kavecl}). This is done by explicitly performing the integral by using contour integration (see \ref{PSD2contint}) resulting in
\begin{equation}
\left< x^4 \right> = \frac{3 \pi^2 \bar{S}_{F^2 F^2}^{\rm th}}{\Gamma \omega_0^2 m^4 \left( 4 \omega_0^2 + 3 \Gamma^2 \right)}.
\label{x4aveclres}
\end{equation}

\noindent Combining this with Eq.~(\ref{x4avedefcl}), we can solve for $\bar{S}_{F^2 F^2}^{\rm th}$ for which we find
\begin{equation}
\bar{S}_{F^2 F^2}^{\rm th} = \frac{\Gamma m^2 \left( 4 \omega_0^2 + 3 \Gamma^2 \right) \left( k_B T \right)^2}{\pi^2 \omega_0^2}.
\label{SthF2F2}
\end{equation}
\noindent Putting this result into Eq.~(\ref{PSD2unnormcl}) we get the final form for the second-order PSD given by
\begin{equation}
\bar{S}_{x^2 x^2}(\omega) = \frac{16 \Gamma \left( 4 \omega_0^2 + 3 \Gamma^2 \right) \left( k_B T \right)^2}{m^2 \omega_0^2 \left( \omega^2 + \Gamma^2 \right) \left( \left( \omega^2 - 4 \omega_0^2 \right)^2 + 4 \omega^2 \Gamma^2 \right)}.
\label{PSD2fincl}
\end{equation}
\noindent This equation can be simplified if we consider the high-$Q$ limit, in which case we can express our PSD as

\begin{equation}
\bar{S}_{x^2 x^2}(\omega) = \frac{64 \Gamma \left( k_B T \right)^2}{m^2 \left( \omega^2 + \Gamma^2 \right) \left( \left( \omega^2 - 4 \omega_0^2 \right)^2 + 4 \omega^2 \Gamma^2 \right)}.
\label{PSD2highQcl}
\end{equation}

As we did above, we will now investigate the maximum values of the second-order PSD. The second-order PSD has three peaks, corresponding to three local maxima. Beginning by evaluating the peak at $\omega=0$, we find
\begin{equation}
\bar{S}_{x^2 x^2}^{\rm DC} = \bar{S}_{x^2 x^2}(0) = \frac{\left( 4 \omega_0^2 + 3 \Gamma^2 \right) \left( k_B T \right)^2}{\Gamma m^2 \omega_0^6} \approx \frac{4}{\Gamma} \left( \frac{k_B T}{m \omega_0^2} \right)^2,
\label{PSD2maxDC}
\end{equation}
\noindent where in the last step we have taken the high-$Q$ approximation.

\begin{figure}[h!]
\includegraphics[width = \columnwidth]{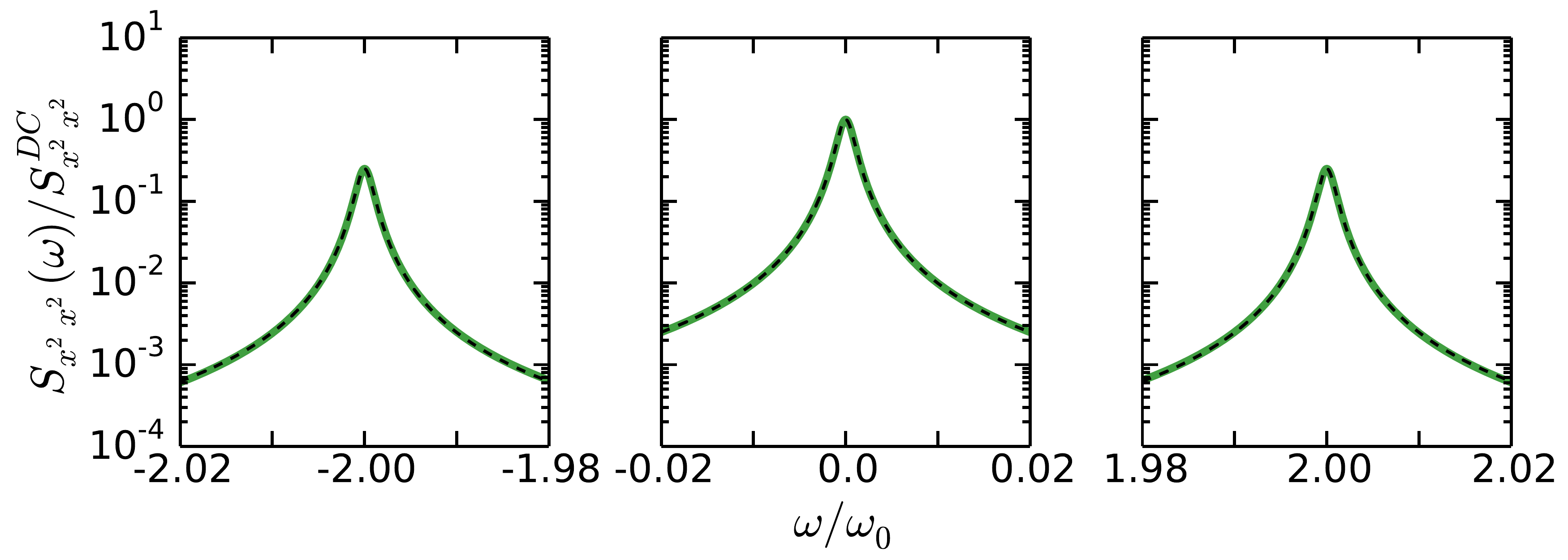}
\caption{The second order classical PSD (solid green) in Eq.~(\ref{PSD2fincl}) and its quantum counterpart (dashed black) given in Eq.~(\ref{PSD2dampfinqu}) in the regions surrounding the peaks at $\omega = 0,\pm 2\omega_0$. Here we have used $\braket{n}=1000$ and $Q=1000$ such that the plots are presented in a frequency window of 10$\Gamma$ on each side of the peak as before. These PSDs are normalized such that their DC peak value is 1.}
\label{2ndOrderQMCL}
\end{figure}

We now perform the same calculation for $\omega = \pm 2 \omega_0$. Fortunately, due to the symmetry of the second-order classical PSD, both of these peaks will have the same maximum value, just as the $\omega = \pm \omega_0$ peaks did in the first-order case. Evaluating the PSD at these two resonant frequencies, we find
\begin{equation}
\bar{S}_{x^2 x^2}^{2 \omega_0} = \bar{S}_{x^2 x^2}(\pm 2 \omega_0) = \frac{\left( 4 \omega_0^2 + 3 \Gamma^2 \right) \left( k_B T \right)^2}{\Gamma m^2 \omega_0^4 \left( 4 \omega_0^2 + \Gamma^2 \right)} \approx \frac{1}{\Gamma} \left( \frac{k_B T}{m \omega_0^2} \right)^2,
\label{PSD2max2w0}
\end{equation}
\noindent where we have again made the high-$Q$ approximation for the last step. Comparing these three maxima (see Fig.~\ref{2ndOrderQMCL}), we find that there is a global maximum at $\omega = 0$, with two local maxima at $\omega = \pm 2 \omega_0$. As well, we find that for high-$Q$, the maxima are related by $\bar{S}_{x^2 x^2}^{\rm DC} = 4 \bar{S}_{x^2 x^2}^{2 \omega_0}$.

We can also use these maxima to determine the FWHM at each of the peaks. Here we use a different approach than we did for the first-order case, due to the differing peak heights, as well as the fact that the denominator depends on the frequency to the sixth power. To simplify our calculations, we assume the high-$Q$ limit to begin with and expand about a small deviation from the resonance frequency $\delta \omega$, which at the half maximum we assume to be on the order of $\Gamma$.

We begin by recalculating the width of the peaks in the first-order case to verify the effectiveness of this method. To do this we evaluate the first-order PSD at $\omega = \omega_0 + \delta \omega$ and equate it to half of its maximum value, which is given in Eq.~(\ref{PSD1max}), resulting in $\bar{S}_{xx} (\omega_0 + \delta \omega) = k_B T / m \omega_0^2 \Gamma$. Evaluating this expression gives
\begin{equation}
\begin{split}
&\left[ \left( \omega_0 + \delta \omega \right)^2 - \omega_0^2 \right]^2 + \left( \omega_0 + \delta \omega \right)^2 \Gamma^2 - 2 \Gamma^2 \omega_0^2 \\
&\approx 4 \omega_0^2 (\delta \omega)^2 - \omega_0^2 \Gamma^2 = 0, \\
&\Rightarrow \delta \omega = \pm \frac{\Gamma}{2},
\label{PSD1halfw}
\end{split}
\end{equation}
\noindent where going from the first step to the second step we have neglected all terms higher than second-order in $\delta \omega \sim \Gamma$. From this result, we find the FWHM as $\Delta \omega = 2 | \delta \omega | = \Gamma$ as we found above, verifying that in the high-$Q$ case, the two methods give the same result. By using this approach, we have implicitly assumed that we are only dealing with the $\omega = \omega_0$ peak, while ignoring the $\omega = -\omega_0$ peak. This is due to the fact that in this analysis, we are only concerned with frequencies resulting from a small expansion around the peak of interest. However, this is inconsequential as the symmetry of the PSD ensures the negative frequency peak will have the same result.

Now that we have verified the efficacy of this method, we apply it to the second-order PSD at $\omega = 0$. Evaluating our PSD at $\delta \omega$ and equating it to the half maximum at the DC peak we find
\begin{equation}
\begin{split}
&\left[ (\delta \omega)^2 + \Gamma^2 \right] \left[ \left( (\delta \omega)^2 - 4 \omega_0^2 \right)^2 + 4 (\delta \omega)^2 \Gamma^2 \right] - 32 \Gamma^2 \omega_0^4 \\
&\approx 16 \omega_0^4 (\delta \omega)^2 - 16 \omega_0^4 \Gamma^2 = 0, \\
&\Rightarrow \delta \omega = \pm \Gamma.
\label{PSD2halfwDC}
\end{split}
\end{equation}
\noindent From this result, we find the FWHM at the DC peak to be $\Delta \omega_{\rm DC} = 2 |\delta \omega | = 2 \Gamma$, which is twice the value of the peaks in the first-order case.

We now perform the same analysis for the peak at $\omega = 2 \omega_0$ (as usual symmetry ensures the same results at $\omega = -2 \omega_0$). Here we now expand about $2 \omega_0$, such that we evaluate our PSD's half maximum at $\omega = 2 \omega_0 + \delta \omega$, resulting in the expression
\begin{equation}
\begin{split}
&\left[ (2\omega_0 + \delta \omega)^2 + \Gamma^2 \right] \Big[ \left( (2 \omega_0 + \delta \omega)^2 - 4 \omega_0^2 \right)^2 \\ &+ 4 (2\omega_0 + \delta \omega)^2 \Gamma^2 \Big] - 32 \Gamma^2 \omega_0^2 \left( 4 \omega_0^2 + \Gamma^2 \right) \\
&\approx 64 \omega_0^4 (\delta \omega)^2 - 64 \omega_0^4 \Gamma^2 = 0, \\
&\Rightarrow \delta \omega = \pm \Gamma.
\label{PSD2halfw2w0}
\end{split}
\end{equation}
\noindent Therefore, the FWHM of the peaks at $\omega = \pm 2 \omega_0$ is $\Delta \omega_{2 \omega_0} = 2 | \delta \omega | = 2 \Gamma$, which is identical to what we found for the DC peak.

\begin{figure}[h!]
\includegraphics[width = \columnwidth]{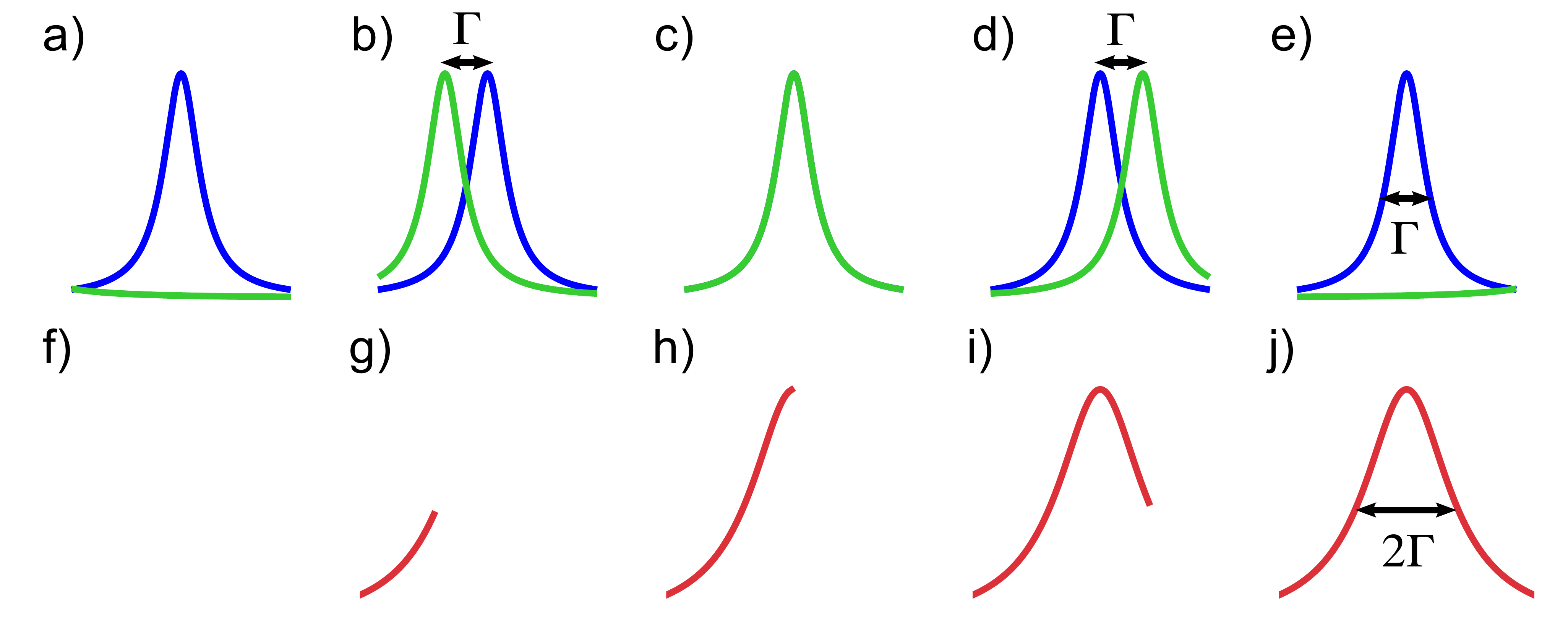}
\caption{A demonstration of how a convolution of two peaked functions with a width of $\Gamma$ produces a peak with a width of $2\Gamma$. In panels a)-e) the green curve is swept across the stationary blue curve while the red curve in f)-j) maps out the shared overlap area beneath the two functions, which is how we can conceptually think of convolution. In panel a) there is little overlap between the blue and green curve, so our curve in f) does not yet exist. In b), the peak of the green function has approached to within $\Gamma$ of the blue. In this case, the overlap causes half of the green or blue curve to be integrated, such that in g) the red curve is at its half-maximum value. In c), when the blue and green functions are perfectly overlapping, the red curve in h) is at its maximum value. In d) we have the opposite case to b), where the green curve is now receding and the red curve in i) is decreasing through its half-maximum. Finally in e), the green curve has left the frame and the majority of the overlap with the blue function has occurred. The end result is the red function in j) with twice the width of the two peaks that were convolved to produce it. As a final note, we mention that this process can be performed any number of times, convolving $k$ curves of width $\Gamma$ a total of $k-1$ times to produce a final distribution with width $k \Gamma$. Thus we have justified our choice of half-width $k \Gamma / 2$ in Eq.~(\ref{deltak}).}
\label{ConvFig_Label}
\end{figure}

It is interesting to note that the second-order PSD has peaks with twice the width of the first-order PSD. We will briefly investigate what this means physically. For the first-order PSD, the FWHM gives a measure of the spread of frequencies around resonance that a single quantum will have. The larger the width, the larger this spread. Now for the second-order PSD, we are looking at processes involving two quanta. Therefore, the spread of accessible frequencies doubles, as we combine the frequency distributions of each. This becomes mathematically apparent by investigating Eq.~(\ref{PSD2defcl}), as the overlap of the mechanical susceptibility with itself in the convolution integral reaches its half-maximum value when each first-order peak is $\sim \Gamma$ away from the other,  producing a second-order function with peaks of twice the width. See Fig.~\ref{ConvFig_Label} for a more detailed explanation of this effect.

\subsection{Quantum}
\label{PSD2qu}

We now look to calculate the second-order PSD in the quantum case. To do this, we follow a methodology similar to what we used to calculate the first-order quantum PSD by first finding the second-order ACF and using it to find the corresponding PSD. Starting by determining the second-order ACF, we take $k=2$ in Eq.~(\ref{ACFkdefqu2}) to obtain
\begin{equation}
\begin{split}
&R_{x^2 x^2}(t) = \left< \hat{x}^2(t) \hat{x}^2 (0) \right> \\ 
&= x_{\rm zpf}^4 \left( 2 \left[ \left( \braket{n} + 1 \right) e^{-i\omega_0 t} + \braket{n} e^{i\omega_0 t} \right]^2 + \left[ 2 \braket{n}^2 +1 \right]^2 \right) \\
&= x_{\rm zpf}^4 \Big[2 \left( \braket{n} + 1 \right)^2 e^{-2i\omega_0 t} + 2 \braket{n}^2 e^{2i\omega_0 t} \\
&+ 8 \braket{n} \left(\braket{n} + 1 \right)  + 1 \Big].
\label{ACF2defqu}
\end{split}
\end{equation}
\noindent By Fourier transforming the above equation we obtain the second-order PSD of the quantum harmonic oscillator 
\begin{equation}
\begin{split}
&S_{x^2 x^2}(\omega) = 2\pi x_{\rm zpf}^4 \Big[ 2 \left( \braket{n} + 1 \right)^2 \delta(\omega - 2 \omega_0) \\
&+ 2 \braket{n}^2 \delta(\omega + 2 \omega_0) + \left( 8 \braket{n} \left( \braket{n} + 1 \right) + 1 \right) \delta(\omega) \Big],
\label{PSD2finqu}
\end{split}
\end{equation}
\noindent which could also have been obtained by taking $k=2$ in Eq.~(\ref{PSDkdefqu}). This result agrees with that found in Eq.~(S3) of \cite{kaviani} and is also verified using the fluctuation-dissipation theorem (see \ref{PSDFDT}).

\begin{figure}[h!]
\includegraphics[width = \columnwidth]{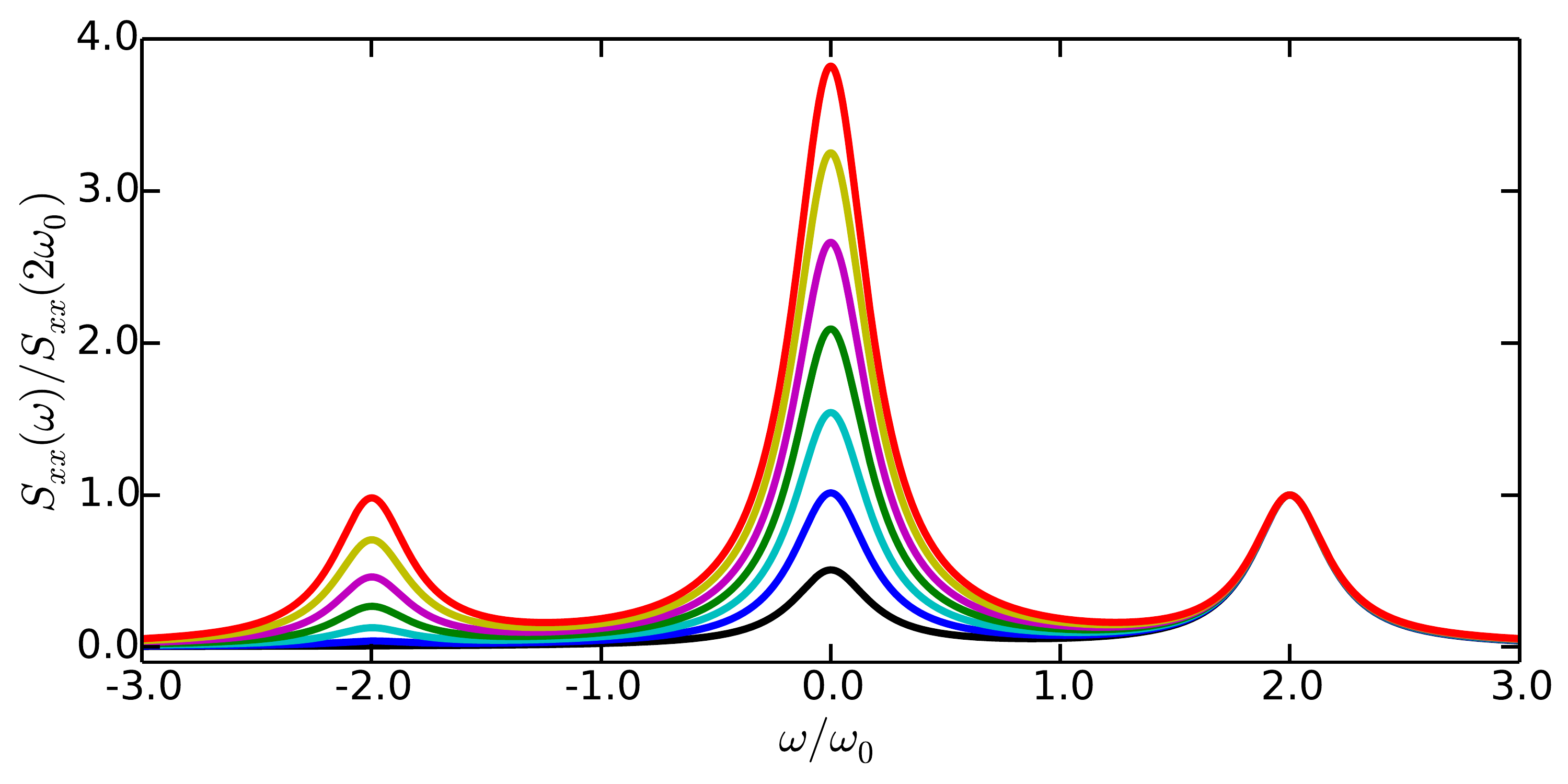}
\caption{The second-order quantum PSD, with each trace normalized such that the peak at $\omega = 2 \omega_0$ is 1, vs frequency in terms of the resonant frequency. The color scheme is the same as in Fig.~\ref{1stOrderAsymm} so that $\braket{n}$ = 0, 0.2, 0.5, 1, 2, 5, 100, as we move up in colors and a quality factor of 5 is chosen for clarity. In this case, the relative height of the peak at $\omega = 2 \omega_0$ stays constant, while the other two peaks decrease as $\braket{n}$ is reduced. Notice that in this case, when $\braket{n}=0$, the $\omega = -2\omega_0$ peak vanishes, while the DC peak still has a finite height, albeit significantly reduced.}
\label{2ndOrderAsymm}
\end{figure}

Eq.~(\ref{PSD2finqu}) also exhibits asymmetric qualities, similar to the first-order case, which are displayed in Fig.~\ref{2ndOrderAsymm}. As we would expect from the classical case, this function contains peaks at $\omega = 0, \pm 2 \omega_0$. Furthermore, by inspecting the coefficients of $\braket{n}^2$ in each term, we find $S_{x^2 x^2}^{\rm DC} = 4 S_{x^2 x^2}^{2 \omega_0}$ for the large $\braket{n}$ case associated with the classical regime, agreeing with what we found in Section \ref{PSD2cl}.

Assigning a physical interpretation to the above peaks, two quanta are annihilated at the $\omega = - 2 \omega_0$ peak, while we have creation of two quanta at the $\omega = 2 \omega_0$ peak. However, we now have a new DC term at $\omega = 0$. Physically, this peak corresponds to the second-order process by which either a quantum is created then annihilated, or annihilated then created, with no net change to the system. These processes are unique to a nonlinear system, as we require a two-step procedure, which is prohibited for a linear system. In cavity optomechanics, this term causes a DC shift in the optical cavity's resonance frequency in proportion to the number of phonons in the mechanical resonator, providing an avenue by which we can perform a QND measurement of these quanta \cite{gangat}. In addition to this effect, cavity optomechanical systems also exhibit other second-order effects corresponding to two-phonon processes, leading to phenomena such as mechanical cooling/squeezing \cite{nunnenkamp}, as well as optomechanically induced transparency \cite{huang,borkje}.

As we did in the first-order case, it is interesting to investigate the $T=0$ limit of the second-order quantum PSD. Taking $\left< n \right> = 0$ we are left with
\begin{equation}
S^0_{x^2 x^2}(\omega) = 2 \pi x_{\rm zpf}^4 [ 2  \delta(\omega - 2 \omega_0) + \delta(\omega) ].
\label{PSD2T0}
\end{equation}
\noindent Again, processes involving the initial annihilation of quanta vanish at $T=0$, leaving the two remaining peaks at $\omega = 2 \omega_0$ and $\omega = 0$, corresponding to the creation of two quanta and the creation of a single quantum followed immediately by its annihilation. What is surprising is that the relative height of the DC peak and the $2 \omega_0$ peak has decreased eight-fold, as the DC peak is now half of that at $2 \omega_0$.

We can also determine the second-order PSD of the damped harmonic oscillator by taking $k=2$ in Eq.~(\ref{PSDkdampqu}) for which the result is
\begin{equation}
\begin{split}
\tilde{S}_{x^2 x^2}(\omega) = 2 \Gamma x_{\rm zpf}^4 &\Bigg[ \frac{2 \left( \braket{n} + 1 \right)^2}{\left( \omega - 2 \omega_0 \right)^2 + \Gamma^2} + \frac{2 \braket{n}^2}{ \left( \omega + 2 \omega_0 \right)^2 + \Gamma^2} \\ 
&+ \frac{8 \braket{n} \left( \braket{n} + 1 \right) + 1}{\omega^2 + \Gamma^2} \Bigg].
\label{PSD2dampfinqu}
\end{split}
\end{equation}
\noindent The zero temperature limit of the above equation can be obtained by taking $\braket{n} = 0$ as usual, resulting in
\begin{equation}
\tilde{S}^0_{x^2 x^2}(\omega) = 2 \Gamma x_{\rm zpf}^4 \left[ \frac{2}{\left( \omega - 2 \omega_0 \right)^2 + \Gamma^2} + \frac{1}{\omega^2 + \Gamma^2} \right].
\label{PSD2dampT0}
\end{equation}

It is also interesting to investigate the thermal average $\left< \hat{x}^4 \right>$ for all $T$. Inputting either the undamped PSD of Eq.~(\ref{PSD2finqu}) or the damped PSD of Eq.~(\ref{PSD2dampfinqu})  into Eq.~(\ref{a2defPSDqu}) as we did in the first-order case we obtain
\begin{equation}
\left< \hat{x}^4 \right> =  3 x^4_{\rm zpf} \left( 2 \braket{n} + 1 \right)^2,
\label{x4avequ}
\end{equation}
\noindent consistent with the $k=2$ case of Eq.~(\ref{x2kavequ}). For $T=0$, this reduces to $\braket{\hat{x}^4} = 3 x_{\rm zpf}^4$, which is in direct correspondence to the classical case given in Eq.~(\ref{x4avedefcl}), where we have replaced the thermal drive with a quantum one.

\subsection{Classical Correspondence}
\label{PSD2cc}

We conclude this section by ensuring that our second-order quantum PSD obeys the correspondence principle. Beginning with Eq.~(\ref{PSD2dampfinqu}) and using the same approximations as we did for the first-order PSD, we find (see \ref{ccapp})
\begin{equation}
\begin{split}
&\tilde{S}_{x^2 x^2}(\omega) \approx \Gamma \left( \frac{k_B T}{m \omega_0^2} \right)^2 \\
&\times \bigg[ \frac{1}{(\omega - 2 \omega_0)^2 +  \Gamma^2} + \frac{1}{(\omega + 2 \omega_0)^2 + \Gamma^2} + \frac{4}{\omega^2 + \Gamma^2} \bigg] \\
&\approx \frac{64 \Gamma \left( k_B T \right)^2}{m^2 \left( \omega^2 + \Gamma^2 \right) \left( \left( \omega^2 - 4 \omega_0^2 \right)^2 + 4 \omega^2 \Gamma^2 \right)},
\label{PSD2ccapprox}
\end{split}
\end{equation}
\noindent which matches the expression we found in Eq.~(\ref{PSD2highQcl}) such that classical correspondence is again satisfied. This is presented visually in Fig.~\ref{2ndOrderQMCL}.

\section{Conclusion}
\label{conc}

We have presented a method to calculate a general PSD for the classical and quantum harmonic oscillator, corresponding to any power of its position that is a positive integer.  We then investigated the experimentally relevant cases of $k=1$ and $k=2$ associated with the linear and quadratic PSDs, respectively. The expressions for the first-order PSD are well-known \cite{hauer,albrecht,ekinci,krause,clerk2,safavi-naeini2} and are presented here to verify our general model. For the case of the second-order PSD, a number of expressions useful in the field of quantum measurement were calculated, allowing researchers to ascertain whether or not a device is suitable for QND measurements \cite{thompson,gangat,doolin,braginsky,jayich,clerk1,kaviani}. Both of these results were found to agree with an independent determination relying on the fluctuation-dissipation theorem. 

Though this document focused largely on the first- and second-order cases, as these are the regimes that are closely linked with experiment, it is possible higher-order PSDs may become useful in the near future. For instance, quartic or fourth-order coupling has already been achieved \cite{sankey} and is proposed as a method to generate Schr\"odinger cat states \cite{yurke,jacobs}. For this reason, the results presented in this paper provide a useful tool by which a theoretical nonlinear PSD can be determined and fit to experimental harmonic oscillator position spectra of any order. 

\begin{widetext}

\appendix

\section{Definitions}
\label{def}

\subsection{Fourier Transforms}
\label{four}

In this document, we choose to define our Fourier transform for an arbitrary, time-dependent signal $a(t)$ as
\begin{equation}
\mathcal{F} \{ a(t) \}  = \int_{-\infty}^{\infty} \! a(t) e^{i \omega t} \, dt = a(\omega),
\label{fourdef}
\end{equation}
\noindent with the inverse Fourier transform being defined as
\begin{equation}
\mathcal{F}^{-1} \{ a(\omega) \}= \frac{1}{2 \pi} \int_{-\infty}^{\infty} \! a(\omega) e^{-i \omega t} \,  d \omega = a(t).
\label{invfourdef}
\end{equation}
\noindent With these definitions, we find the following useful property of the complex conjugate of the Fourier transform
\begin{equation}
a^*(\omega) = \left(  \int_{-\infty}^{\infty} \! a(t) e^{i \omega t} \, dt \right)^* = \int_{-\infty}^{\infty} \! a(t) e^{-i \omega t} \, dt = a(- \omega).
\label{fourcc}
\end{equation}
\noindent Here we have assumed that the time-varying signal $a(t)$ is real. We also have the expression for the Fourier transform of the $n$th time derivative of $a(t)$ given by
\begin{equation}
\mathcal{F} \left\{ \frac{d^n a}{dt^n} \right\} = \left( -i \omega \right)^n \mathcal{F} \{ a(t) \} = \left( -i \omega \right)^n  a(\omega).
\label{fourderiv}
\end{equation}

\subsection{Convolution}
\label{conv}

For an arbitrary variable $u$, the convolution of two functions $g(u)$ and $h(u)$ is defined as
\begin{equation}
g(u) * h(u) = \int_{-\infty}^{\infty} \! g(u') h(u - u') \, du' ,
\label{convdef}
\end{equation}
\noindent and represents the measure of the area shared by the two functions, as we translate one across the other. A useful property of this operation is that the Fourier transform of a convolution of two signals in the time domain is the product of their Fourier transforms in the frequency domain. That is to say
\begin{equation}
 \mathcal{F} \{ g(t) * h(t) \} = g(\omega) h(\omega),
\label{fourconv}
\end{equation}
\noindent where  $g( \omega ) = \mathcal{F} \{ g(t) \}$ and $h( \omega ) = \mathcal{F} \{ h(t) \}$ are the Fourier transforms of the signals $g(t)$ and $h(t)$. 

We can also use the convolution integral to express the Fourier transform of a product of two functions in the time domain as the convolution of their frequency domain representations. For our Fourier transform definitions, this is given by
\begin{equation}
 \mathcal{F} \{ g(t) h(t) \} = \frac{1}{2 \pi} g(\omega) * h(\omega) .
\label{fourprod}
\end{equation}
\noindent Finally, this property can be extended to a product of $k$ functions in the time domain giving
\begin{equation}
 \mathcal{F} \{ g_1(t) g_2(t) ... g_k(t) \} = \frac{1}{(2 \pi)^{k-1}} g_1(\omega)*g_2(\omega)*...*g_k(\omega),
\label{fourprodk}
\end{equation}
\noindent where the ellipsis (...) is used to indicate that there are $k$ terms in the sequence.

\subsection{Dirac Delta Function}
\label{delta}

Conventionally, the one-dimensional Dirac delta function is defined as
\begin{equation}
\delta (u - u') = \begin{cases} \infty &\mbox{if } u = u', \\ 
0 & \mbox{if } u \ne u', \end{cases} 
\label{deltadef}
\end{equation}
\noindent such that 
\begin{equation}
\int \! f(u') \delta (u-u') \, du'  = f(u),
\label{deltaint}
\end{equation}
\noindent provided that the integral contains $u = u'$ in its range of integration. 

This is not the only way to express the Dirac delta function, however, and here we introduce two alternate definitions which are used above. The first is given by
\begin{equation}
\delta (u - u') = \frac{1}{2 \pi} \int_{-\infty}^{\infty} \! e^{i \nu (u - u')} \, d \nu.
\label{deltafour}
\end{equation}
\noindent This definition proves to be very useful when performing Fourier transform integrals throughout this paper.

The second definition arises from taking the limiting case of a Lorentzian function, such that
\begin{equation}
\delta (u-u') = \lim_{\epsilon \to 0} \frac{1}{\pi} \frac{\epsilon}{(u-u')^2 + \epsilon^2},
\label{deltalor}
\end{equation}
\noindent where $\epsilon$ is the half width at half maximum (HWHM) of the peak. This equation provides one of the simplest ways for introducing width to peaks that are infinitesimally narrow ({\it i.e.} including damping where there was initially none).

\section{Contour Integration}
\label{contint}

Contour integration is a powerful method of integration by which functions containing complex poles can be easily integrated. This technique hinges upon the Cauchy's residue theorem, which states \cite{brown}
\begin{equation}
\oint_{\mathcal{C}} g(z) dz = 2 \pi i \displaystyle \sum^{N}_{l=1} {\rm Res}(g,z_l),
\label{cauchth}
\end{equation}
\noindent where the integral is performed over a positively oriented (counter-clockwise) closed contour $\mathcal{C}$ and $z_l$ are the $N$ poles of $g(z)$ enclosed by $\mathcal{C}$. As well, we have introduced the residue of $g(z)$ at $z_l$ denoted by ${\rm Res}(g,z_l)$. For a function that can be expressed as
\begin{equation}
g(z) = \frac{\theta(z)}{z - z_l},
\label{polefunc}
\end{equation}
\noindent we can determine its residue at $z_l$ as
\begin{equation}
{\rm Res}(g,z_l) = \theta(z_l).
\label{resdef}
\end{equation}

Often, a convenient choice for the contour $\mathcal{C}$ is a semicircle with infinite radius that extends over the region of the complex plane with ${\rm Im} \{ z \}> 0$. In this way, the portion of the semicircle that runs along the real axis stretches from $-\infty$ to $\infty$. If we assume that $|g(z)|$ falls off faster than $\frac{1}{z^2}$ as $z \rightarrow \infty$, the portion of the integral performed on the curved part of the contour will be zero, allowing us to write
\begin{equation}
\oint_{\mathcal{C}} \! g(z) \, dz = \int_{-\infty}^{\infty} \! g(z) \, dz = 2 \pi i \displaystyle \sum^{N}_{l=1} {\rm Res}(g,z_l).
\label{contintfin}
\end{equation}
\noindent Our choice of contour is such that we are now only concerned with the poles in the top half of the complex plane. We could have alternatively chosen the semicircle enclosing the poles in the bottom half of the complex plane, provided we account for the minus signs that will arise due to our differing contour orientation. The integral given in Eq.~(\ref{contintfin}) is useful in computing a number of quantities for PSDs.

\section{Thermal Average of ${\bf x^{2k}}$ for the Classical Harmonic Oscillator}
\label{x2kave}

In order to normalize the PSDs calculated in this document, it is important to know the thermal average of the position to even powers, that is $\left< x^{2k} \right>$. For a one-dimensional, classical system, the thermal average of a quantity $A$ is given by
\begin{equation}
\left< A \right> = \frac{ \displaystyle \int \hspace{-4pt} \int \! A e^{-\beta H} \, dx \, dp}{\displaystyle \int \hspace{-4pt} \int \! e^{-\beta H} \,  dx \, dp},
\label{Aavecl}
\end{equation}
\noindent where $\beta = 1 / k_B T$ and $H = H(x,p)$ is the Hamiltonian of the system as a function of position $x$ and momentum $p$ and the integrals are performed over the entire region of each corresponding phase space.

In order to calculate the thermal average of  $x^{2k}(t)$ for the one-dimensional harmonic oscillator, we input the Hamiltonian from Eq.~(\ref{CHOenergy}) into Eq.~(\ref{Aavecl}) with $A=x^{2k}$ resulting in
\begin{equation}
\left< x^{2k} \right> = \frac{\displaystyle \int_{-\infty}^{\infty} \! x^{2k} e^{- \frac{1}{2} \beta m \omega_0^2 x^2} \, dx \int_{-\infty}^{\infty} \! e^{- \frac{\beta p^2}{2m}} \, dp }{\displaystyle \int_{-\infty}^{\infty} \! e^{- \frac{1}{2} \beta m \omega_0^2 x^2} \, dx \int_{-\infty}^{\infty} \! e^{- \frac{\beta p^2}{2m}} \, dp } = \frac{\displaystyle \int_{-\infty}^{\infty} \! x^{2k} e^{- \frac{1}{2} \beta m \omega_0^2 x^2} \, dx}{\displaystyle \int_{-\infty}^{\infty} \! e^{- \frac{1}{2} \beta m \omega_0^2 x^2} \, dx}.
\label{x2kaveclint}
\end{equation}
\noindent To evaluate the second line of Eq.~(\ref{x2kaveclint}), we calculate the integral in the numerator, which can be shown (by induction) to be
\begin{equation}
\int_{-\infty}^{\infty} \! x^{2k} e^{- \frac{1}{2} \beta m \omega_0^2 x^2} \, dx = \frac{(2k)!}{(2 \beta m \omega_0^2)^k k!} \int_{-\infty}^{\infty} \! e^{- \frac{1}{2} \beta m \omega_0^2 x^2} \, dx.
\label{x2kinttox0int}
\end{equation}
\noindent We can therefore input Eq.~(\ref{x2kinttox0int}) into Eq.~(\ref{x2kaveclint}) to obtain
\begin{equation}
\left< x^{2k} \right> = \frac{(2k)!}{(2 \beta m \omega_0^2)^k k!} = x_{\rm th}^{2k} \frac{(2k)!}{2^k k!},
\label{x2kclapp}
\end{equation}
\noindent where $x_{\rm th} = \sqrt{1/ \beta m \omega_0^2} = \sqrt{k_B T / m \omega_0^2}$ is the root-mean-square amplitude of our classical thermally driven motion. Inputting $k=1$ and $k=2$ we obtain
\begin{equation}
\left< x^2 \right> = \frac{1}{\beta m \omega_0^2} = \frac{k_BT}{m \omega_0^2} = x_{\rm th}^2,
\label{x2aveapp}
\end{equation}
\noindent and
\begin{equation}
\left< x^4 \right> = \frac{3}{(\beta m \omega_0^2)^2} = 3 \left( \frac{k_B T}{m \omega_0^2} \right)^2 = 3 x_{\rm th}^4,
\label{x4aveapp}
\end{equation}
\noindent which are of special interest for this document as they are required to properly normalize the first- and second-order classical PSDs.

\section{Wick's Theorem}
\label{Wick}

Wick's theorem \cite{mahan,fetter} is a powerful operator identity that is often used in quantum field theory to simplify the products of creation and annihilation operators. Here we show how it can be used in the context of determining the $k$th-order ACF for the position operator of the quantum harmonic oscillator. 

In order to properly introduce Wick's theorem, we must first define a number of ordering operations which can be performed on products of quantum mechanical operators. The first such operation is known as time-ordering and will be enforced using the time-ordering operator $\mathcal{T}$. When applied to a product of time-dependent operators, time-ordering ensures that operators with a larger time argument appear left of those with a smaller one. For instance, when applied to the product of bosonic operators $\hat{x}(t_i) \hat{x}(t_j)$, the time-ordering operator produces
\begin{equation} 
\mathcal{T}\{\hat{x}(t_i) \hat{x}(t_j)\} = \begin{cases} \hat{x}(t_i) \hat{x}(t_j) &\mbox{for $t_i > t_j$}, \\ 
\hat{x}(t_j) \hat{x}(t_i) & \mbox{for $t_i < t_j$}. \end{cases} 
\label{timeord}
\end{equation}
\noindent We can also introduce an anti-time-ordering operator $\bar{\mathcal{T}}$, which has the opposite effect of the time-ordering operator. That is to say the anti-time-ordered product is organized from left to right by ascending time arguments. Applying the anti-time-ordering operator to the example given in Eq.~(\ref{timeord}) now instead gives
\begin{equation} 
\bar{\mathcal{T}}\{\hat{x}(t_i) \hat{x}(t_j)\} = \begin{cases} \hat{x}(t_j) \hat{x}(t_i) &\mbox{for $t_i > t_j$}, \\ 
\hat{x}(t_i) \hat{x}(t_j) & \mbox{for $t_i < t_j$}. \end{cases} 
\label{antitimeord}
\end{equation}

Next, we introduce the concept of normal ordering. This operation is performed by the normal ordering operator $\mathcal{N}$, which takes an arbitarily ordered product of creation and annihilation operators and arranges them so that all of the creation operators are on the left of the annihiliation operators. A simple example of this process is the normal ordering of the product $\hat{b} \hat{b}^\dag \hat{b} \hat{b}^\dag$, which is given by
\begin{equation} 
\mathcal{N} \{ \hat{b} \hat{b}^\dag \hat{b} \hat{b}^\dag \} =  \hat{b}^\dag \hat{b}^\dag \hat{b} \hat{b}.
\label{normord}
\end{equation}
\noindent This operator ordering has the very useful property that when acting on the ground state, it always produces zero as a result of the rightmost annihilation operator.

Finally, we introduce the contraction of two operators, which for the case of two position operators is given by
\begin{equation} 
\contraction{}{\hat{x}}{(t_i) }{\hat{x}}
\hat{x}(t_i) \hat{x}(t_j) = \mathcal{T} \{ \hat{x}(t_i) \hat{x}(t_j) \} - \mathcal{N} \{ \hat{x}(t_i) \hat{x}(t_j) \}.
\label{contract}
\end{equation}
\noindent Using the expression for $\hat{x}(t)$ given in Eq.~(\ref{xopt}), we can explicitly calculate the contraction above as
\begin{equation} 
\contraction{}{\hat{x}}{(t_i) }{\hat{x}}
\hat{x}(t_i) \hat{x}(t_j) = x_{\rm zpf}^2 \times \begin{cases} e^{-i\omega_0 ( t_i - t_j)} &\mbox{for $t_i > t_j$}, \\ 
  e^{-i\omega_0 ( t_j - t_i)}& \mbox{for $t_i < t_j$}. \end{cases} 
\label{contractex}
\end{equation}
\noindent where we have used the commutation relation $[\hat{b},\hat{b}^\dag]=1$. It is interesting to note that the contraction shown here is simply a complex number (not an operator) independent of the number of quanta in the system.

We now have the machinery required to properly present Wick's theorem, which states that a time-ordered product of operators comprised of creation and annihilation operators can be represented as the normal ordering of the product, plus a sum over all possible unique contractions of the operators within the product. For the position operators of the quantum harmonic oscillator, this is mathematically expressed as \cite{fetter}
\begin{equation}
\begin{split}
&\mathcal{T} \{ \hat{x}(t_N) \hat{x}(t_{N-1}) ... \hat{x}(t_2) \hat{x}(t_1) \} = \mathcal{N} \{ \hat{x}(t_N) \hat{x}(t_{N-1}) ... \hat{x}(t_2) \hat{x}(t_1) \} + \sum_{\rm single} \mathcal{N} \{ \hat{x}(t_N) \hat{x}(t_{N-1}) ... \hat{x}(t_2) \hat{x}(t_1) \} \\
&+ \sum_{\rm double} \mathcal{N} \{ \hat{x}(t_N) \hat{x}(t_{N-1}) ... \hat{x}(t_2) \hat{x}(t_1) \} + ... + \sum_{\rm all} \mathcal{N} \{ \hat{x}(t_N) \hat{x}(t_{N-1}) ... \hat{x}(t_2) \hat{x}(t_1) \},
\label{Wickeq}
\end{split}
\end{equation}
\noindent where the sums are carried out by performing all of the specified unique contractions (single, double, triple, etc.)~and $N=2k$ is the number of position operators we are considering.

We now look at how this theorem can be used to reduce the correlation functions comprised of $2k$ operators (known as $2k$-point correlators) used to calculate the $k$th-order ACFs to a sum of products of two-point correlators. We begin by introducing the following useful identity comprised of correlators of normal ordered position operators
\begin{equation}
\begin{split}
\langle \mathcal{N} \{ \hat{x}(t_N) \hat{x}(t_{N-1}) ... \hat{x}(t_2) \hat{x}(t_1) \} \rangle = \sum_{\rm TO} \langle \mathcal{N} \{ \hat{x}(t_{i_k}) \hat{x}(t_{j_k}) \} \rangle ... \langle \mathcal{N} \{ \hat{x}(t_{i_1}) \hat{x}(t_{j_1}) \} \rangle,
\label{normcorr}
\end{split}
\end{equation}
\noindent where the $\displaystyle \sum_{\rm TO}$ indicates a sum over all unique products of time-ordered two-point correlators on the RHS of the equation, that is $t_i > t_j$ for all sub-indices on $i$ and $j$ and there are no repeated terms. We also note that this identity hinges on the fact that we can write $\langle ( \hat{b}^\dag)^k \hat{b}^k \rangle = k! \braket{\hat{b}^\dag \hat{b}}^k$.

Next we rearrange  Eq.~(\ref{contract}) and take the thermal average resulting in
\begin{equation}
\begin{split}
\contraction{\langle \mathcal{T} \{ \hat{x}(t_i) \hat{x}(t_j) \} \rangle = \langle \mathcal{N} \{ \hat{x}(t_i) \hat{x}(t_j) \}  \rangle + \langle}{\hat{x}}{(t_i) }{\hat{x}}
\langle \mathcal{T} \{ \hat{x}(t_i) \hat{x}(t_j) \} \rangle  &= \langle \mathcal{N} \{ \hat{x}(t_i) \hat{x}(t_j) \}  \rangle + \langle \hat{x}(t_i) \hat{x}(t_j) \rangle.
\label{to2pcorr}
\end{split}
\end{equation}
\noindent We can now imagine taking the product of $k$ of these correlators which results in
\begin{equation}
\begin{split}
&\langle \hat{x}(t_N) \hat{x}(t_{N-1})\rangle ... \langle \hat{x}(t_2) \hat{x}(t_1)\rangle \rangle = \langle \mathcal{N} \{ \hat{x}(t_N) \hat{x}(t_{N-1}) \}  \rangle ... \langle \mathcal{N} \{ \hat{x}(t_2) \hat{x}(t_1) \}  \rangle 
\contraction{+ \sum_i \langle}{\hat{x}}{(t_i) }{\hat{x}}
+ \sum_i \langle \hat{x}(t_i) \hat{x}(t_{i-1}) \rangle \prod_{l \ne i} \langle \mathcal{N} \{ \hat{x}(t_l) \hat{x}(t_{l-1}) \}  \rangle \\
\contraction{+ \sum_{i,j} \langle}{\hat{x}}{(t_i) }{\hat{x}}
\contraction{+ \sum_{i,j} \langle \hat{x}(t_i) \hat{x}(t_{i-1}) \rangle \langle}{\hat{x}}{(t_j) }{\hat{x}}
&+ \sum_{i,j} \langle \hat{x}(t_i) \hat{x}(t_{i-1}) \rangle \langle \hat{x}(t_j) \hat{x}(t_{j-1}) \rangle \prod_{l \ne i,j} \langle \mathcal{N} \{ \hat{x}(t_l) \hat{x}(t_{l-1}) \}  \rangle + ... 
\contraction{+ \langle}{\hat{x}}{(t_N) }{\hat{x}}
\contraction{+ \langle \hat{x}(t_N) \hat{x}(t_{N-1}) \rangle ... \langle }{\hat{x}}{(t_2) }{\hat{x}}
+ \langle \hat{x}(t_N) \hat{x}(t_{N-1}) \rangle ... \langle \hat{x}(t_2) \hat{x}(t_1) \rangle,
\label{k2pcorr}
\end{split}
\end{equation}
\noindent where we have taken $t_N > t_{N-1} > ... > t_2 > t_1$ such that we can drop the time-ordering operators on the LHS.  In other words, the product on the LHS is equal to a product of $k$ normal ordered two-point correlators, plus terms containing single contractions, plus terms containing double contractions, etc., all the way until we reach the term containing $k$ contractions. Finally, we sum over all unique combinations of Eq.~(\ref{k2pcorr}) such that each two-point correlator in the product on the LHS is time-ordered to obtain
\begin{equation}
\begin{split}
&\sum_{\rm TO} \langle \hat{x}(t_{i_k}) \hat{x}(t_{j_k})\rangle ... \langle \hat{x}(t_{i_1}) \hat{x}(t_{j_1})\rangle \\
 &= \sum_{\rm TO} \langle \mathcal{N} \{ \hat{x}(t_{i_k}) \hat{x}(t_{j_k}) \}  \rangle ... \langle \mathcal{N} \{ \hat{x}(t_{i_1}) \hat{x}(t_{j_1}) \}  \rangle 
\contraction{+ \sum_{\rm TO} \sum_i \langle}{\hat{x}}{(t_{i_m})}{\hat{x}}
+ \sum_{\rm TO} \sum_m \langle \hat{x}(t_{i_m}) \hat{x}(t_{j_m}) \rangle \prod_{l \ne m} \langle \mathcal{N} \{ \hat{x}(t_{i_l}) \hat{x}(t_{j_l}) \}  \rangle \\
\contraction{+ \sum_{\rm TO} \sum_{i,j} \langle}{\hat{x}}{(t_{i_m})}{\hat{x}}
\contraction{+ \sum_{\rm TO} \sum_{m,n} \langle \hat{x}(t_{i_m}) \hat{x}(t_{j_m}) \rangle \langle}{\hat{x}}{(t_{i_n})}{\hat{x}}
&+ \sum_{\rm TO} \sum_{m,n} \langle \hat{x}(t_{i_m}) \hat{x}(t_{j_m}) \rangle \langle \hat{x}(t_{i_n}) \hat{x}(t_{j_n}) \rangle \prod_{l \ne m,n} \langle \mathcal{N} \{ \hat{x}(t_l) \hat{x}(t_{l-1}) \}  \rangle + ... + 
\contraction{+ \sum_{\rm TO} \langle}{\hat{x}}{(t_{i_k}) }{\hat{x}}
\contraction{+ \sum_{\rm TO} \langle \hat{x}(t_{i_k}) \hat{x}(t_{j_k}) \rangle ... \langle}{\hat{x}}{(t_{i_1}) }{\hat{x}}
 \sum_{\rm TO} \langle \hat{x}(t_{i_k}) \hat{x}(t_{j_k}) \rangle ... \langle \hat{x}(t_{i_1}) \hat{x}(t_{j_1}) \rangle \\
&= \langle \mathcal{N} \{ \hat{x}(t_N) \hat{x}(t_{N-1}) ... \hat{x}(t_2) \hat{x}(t_1) \} \rangle + \sum_{\rm single} \langle \mathcal{N} \{ \hat{x}(t_N) \hat{x}(t_{N-1}) ... \hat{x}(t_2) \hat{x}(t_1) \} \rangle + \sum_{\rm double} \langle \mathcal{N} \{ \hat{x}(t_N) \hat{x}(t_{N-1}) ... \hat{x}(t_2) \hat{x}(t_1) \} \rangle \\
&+ ... + \sum_{\rm all} \langle \mathcal{N} \{ \hat{x}(t_N) \hat{x}(t_{N-1}) ... \hat{x}(t_2) \hat{x}(t_1) \} \rangle,
\label{k2pcorrsumto}
\end{split}
\end{equation}
\noindent where we have used the identity in Eq.~(\ref{normcorr}) with the same definition of $\displaystyle \sum_{\rm TO}$ as above, as well as the fact that the contraction of two position operators is a complex number independent of the number of quanta in the system such that it can be pulled outside the correlator. By inspecting the final result of Eq.~(\ref{k2pcorrsumto}), we see that by summing over all unique combinations of $k$ time-ordered two-point correlators, we obtain what we would get if we took the thermal average of Eq.~(\ref{Wickeq}). Therefore, this allows us to use Wick's theorem to write a $2k$-point position operator correlator as
\begin{equation}
\begin{split}
&\langle \hat{x}(t_N) \hat{x}(t_{N-1}) ... \hat{x}(t_2) \hat{x}(t_1) \rangle = \sum_{\rm TO} \langle \hat{x}(t_{i_k}) \hat{x}(t_{j_k})\rangle ... \langle \hat{x}(t_{i_1}) \hat{x}(t_{j_1})\rangle.
\label{Wickres}
\end{split}
\end{equation}
\noindent It is this result that allows us to break our $k$th-order ACFs into sums of products of two-point correlators.

Before we go into more detail on how to calculate $k$th-order ACFs using this result, we present a brief example to elucidate the above method by showing how it can be applied to reduce the four-point correlator $\langle \hat{x}(t_4) \hat{x}(t_3) \hat{x}(t_2) \hat{x}(t_1) \rangle$ to a sum of three products of time-ordered two-point correlators of the form $\langle \hat{x}(t_i) \hat{x}(t_j) \rangle$. For this example, we assume that $t_4 > t_3 > t_2 > t_1$ such that we do not need to explicitly write out the time-ordering operator. Using Wick's theorem, we can write this four-point correlator as
\begin{equation}
\begin{split}
\contraction{\langle \hat{x}(t_4) \hat{x}(t_3) \hat{x}(t_2) \hat{x}(t_1) \rangle = \langle \mathcal{N} \{ \hat{x}(t_4) \hat{x}(t_3) \hat{x}(t_2) \hat{x}(t_1) \} \rangle + \langle \mathcal{N} \{ }{\hat{x}}{(t_4)}{\hat{x}}
& \langle \hat{x}(t_4) \hat{x}(t_3) \hat{x}(t_2) \hat{x}(t_1) \rangle = \langle \mathcal{N} \{ \hat{x}(t_4) \hat{x}(t_3) \hat{x}(t_2) \hat{x}(t_1) \} \rangle + \langle \mathcal{N} \{ \hat{x}(t_4) \hat{x}(t_3) \hat{x}(t_2) \hat{x}(t_1) \} \rangle 
\contraction{+ \langle \mathcal{N} \{ }{\hat{x}}{(t_4) \hat{x}(t_3)}{\hat{x}}
\contraction{+ \langle \mathcal{N} \{ \hat{x}(t_4) \hat{x}(t_3) \hat{x}(t_2) \hat{x}(t_1) \} \rangle + \langle \mathcal{N} \{ \hat{x}(t_4) \hat{x}(t_3) \hat{x}(t_2) \hat{x}(t_1) \} \rangle + \langle \mathcal{N} \{ \hat{x}(t_4) }{\hat{x}}{(t_3)}{\hat{x}}
+ \langle \mathcal{N} \{ \hat{x}(t_4) \hat{x}(t_3) \hat{x}(t_2) \hat{x}(t_1) \} \rangle \\ 
\contraction{+ \langle \mathcal{N} \{ }{\hat{x}}{(t_4) \hat{x}(t_3)\hat{x}(t_2)}{\hat{x}}
\contraction{+ \langle \mathcal{N} \{ \hat{x}(t_4) \hat{x}(t_3) \hat{x}(t_2) \hat{x}(t_1) \} \rangle + \langle \mathcal{N} \{ \hat{x}(t_4) }{\hat{x}}{(t_3)}{\hat{x}}
&+ \langle \mathcal{N} \{ \hat{x}(t_4) \hat{x}(t_3) \hat{x}(t_2) \hat{x}(t_1) \} \rangle + \langle \mathcal{N} \{ \hat{x}(t_4) \hat{x}(t_3) \hat{x}(t_2) \hat{x}(t_1) \} \rangle  
\contraction{+\langle \mathcal{N} \{ \hat{x}(t_4) }{\hat{x}}{(t_3)\hat{x}(t_2)}{\hat{x}}
\contraction{+\langle \mathcal{N} \{ \hat{x}(t_4) \hat{x}(t_3) \hat{x}(t_2) \hat{x}(t_1) \} \rangle + \langle \mathcal{N} \{ \hat{x}(t_4) \hat{x}(t_3) }{\hat{x}}{(t_2)}{\hat{x}}
+\langle \mathcal{N} \{ \hat{x}(t_4) \hat{x}(t_3) \hat{x}(t_2) \hat{x}(t_1) \} \rangle + \langle \mathcal{N} \{ \hat{x}(t_4) \hat{x}(t_3) \hat{x}(t_2) \hat{x}(t_1) \} \rangle \\ 
\contraction{+ \langle \mathcal{N} \{}{\hat{x}}{(t_4)}{\hat{x}}
\contraction{+ \langle \mathcal{N} \{ \hat{x}(t_4) \hat{x}(t_3) }{\hat{x}}{(t_2)}{\hat{x}}
&+ \langle \mathcal{N} \{ \hat{x}(t_4) \hat{x}(t_3) \hat{x}(t_2) \hat{x}(t_1) \} \rangle 
\contraction{+ \langle \mathcal{N} \{ }{\hat{x}}{(t_4)\hat{x}(t_3)}{\hat{x}}
\contraction[2ex]{+ \langle \mathcal{N} \{ \hat{x}(t_4) }{\hat{x}}{(t_3)\hat{x}(t_2)}{\hat{x}}
\contraction[2ex]{+ \langle \mathcal{N} \{ \hat{x}(t_4) \hat{x}(t_3) \hat{x}(t_2) \hat{x}(t_1) \} \rangle + \langle \mathcal{N} \{ }{\hat{x}}{(t_4)\hat{x}(t_3)\hat{x}(t_2)}{\hat{x}}
\contraction{+ \langle \mathcal{N} \{ \hat{x}(t_4) \hat{x}(t_3) \hat{x}(t_2) \hat{x}(t_1) \} \rangle + \langle \mathcal{N} \{ \hat{x}(t_4) }{\hat{x}}{(t_3)}{\hat{x}}
+ \langle \mathcal{N} \{ \hat{x}(t_4) \hat{x}(t_3) \hat{x}(t_2) \hat{x}(t_1) \} \rangle + \langle \mathcal{N} \{ \hat{x}(t_4) \hat{x}(t_3) \hat{x}(t_2) \hat{x}(t_1) \} \rangle.
\label{Wick4corr}
\end{split}
\end{equation}
\noindent Alternatively, we could have used Eq.~(\ref{Wickres}) to write our four-point correlator as
\begin{equation}
\langle \hat{x}(t_4) \hat{x}(t_3) \hat{x}(t_2) \hat{x}(t_1) \rangle = \langle \hat{x}(t_4) \hat{x}(t_3) \rangle \langle \hat{x}(t_2) \hat{x}(t_1) \rangle + \langle \hat{x}(t_4) \hat{x}(t_2) \rangle \langle \hat{x}(t_3) \hat{x}(t_1) \rangle + \langle \hat{x}(t_4) \hat{x}(t_1) \rangle \langle \hat{x}(t_3) \hat{x}(t_1) \rangle,
\label{prod4corr}
\end{equation}
\noindent where we have been careful to ensure that each of our two-point correlators are unique and time-ordered. Evaluating this correlator using either Eq.~(\ref{Wick4corr}) or Eq.~(\ref{prod4corr}), we obtain the identical result, namely
\begin{equation}
\begin{split}
& \langle \hat{x}(t_4) \hat{x}(t_3) \hat{x}(t_2) \hat{x}(t_1) \rangle = x_{\rm zpf}^4 \Big[ 2 \braket{n}^2 \Big( e^{-i\omega_0 (t_4 + t_3 - t_2 - t_1)} + e^{i\omega_0 (t_4 + t_3 - t_2 - t_1)} + e^{-i\omega_0 (t_4 - t_3 + t_2 - t_1)} + e^{i\omega_0 (t_4 - t_3 + t_2 - t_1)} \\ 
&+ e^{-i\omega_0 (t_4 - t_3 - t_2 + t_1)} + e^{i\omega_0 (t_4 - t_3 - t_2 + t_1)}  \Big)  + \braket{n} \Big( 4e^{-i\omega_0 (t_4 + t_3 - t_2 - t_1)} + 3e^{-i\omega_0 (t_4 - t_3 + t_2 - t_1)} \\
&+ e^{i\omega_0 (t_4 - t_3 + t_2 - t_1)} + 2e^{-i\omega_0 (t_4 - t_3 - t_2 + t_1)} + 2e^{i\omega_0 (t_4 - t_3 - t_2 + t_1)}  \Big) + e^{-i\omega_0 (t_4 - t_3 + t_2 - t_1)} + 2 e^{-i\omega_0 (t_4 + t_3 - t_2 - t_1)} \Big].
\label{Wick4corrres}
\end{split}
\end{equation}
\noindent We note that the above result is consistent with that found in Eq.~(\ref{ACF2defqu}) for the second order ACF, provided we take $t_4 = t_3 =t$ and $t_2 = t_1 =0$.

Continuing on to calculate the $k$th-order ACF, we now imagine replacing $t_N = t_{N-1} = ... = t_{N-k+1} = t$ and $t_{N-k} = t_{N-k-1} = ... = t_1 = 0$ in Eq.~(\ref{Wickres}), which casts the correlator on the RHS into the form identical to that of Eq.~(\ref{ACFkdefqu1}). At first glance, it would seem that due to the implicit assumption of time-ordering in Eq.~(\ref{Wickres}), this association is only valid for $t > 0$, where we are ensured that our $k$th-order ACF is time-ordered. However, though we have been assuming time-ordering throughout this section, the exact same results can be shown to be true for anti-time-ordering ({\it ie.}~replace all time-ordering operators with anti-time-ordering operators). Therefore, this result is also valid for $t < 0$. The trivial case of $t=0$ can be handled by the continuity between these two domains. Our final result then becomes
\begin{equation}
\begin{split} 
\braket{\hat{x}^k(t) \hat{x}^k(0)} = \displaystyle\sum\limits_{c=0}^{N} A_c\braket{\hat{x}(t) \hat{x}(t)}^c \braket{\hat{x}(t)\hat{x}(0)}^{k - 2c} \braket{\hat{x}(0) \hat{x}(0)}^c,
\label{ACFkWick}
\end{split}
\end{equation}
\noindent where $A_c$ is given by Eq.~(\ref{ACFkAc}) and counts the terms of differing orders in the two-point correlators used here.

\section{Calculation of First-Order Quantum Autocorrelation Functions}
\label{2pACF}

Here we calculate the two-point correlators comprised of $\hat{x}(t)$ and $\hat{x}(0)$ found in Eq.~(\ref{ACFkdefqu1}) which are used as the building blocks to determine any general ACF for the position of the harmonic oscillator to the $k$th power. To calculate these three correlators, we use Eq.~(\ref{xopt}) to obtain
\begin{equation}
\begin{split}
&\braket{\hat{x}(t) \hat{x}(0)} = x_{\rm zpf}^2 \left( \braket{\hat{b} \hat{b}} e^{-i\omega_0 t} + \braket{\hat{b} \hat{b}^{\dag}} e^{-i\omega_0 t} + \braket{\hat{b}^{\dag} \hat{b}} e^{i\omega_0 t} + \braket{\hat{b}^{\dag} \hat{b}^{\dag}} e^{i\omega_0 t} \right), \\
&\braket{\hat{x}(t) \hat{x}(t)} = x_{\rm zpf}^2 \left( \braket{\hat{b} \hat{b}} e^{-2i\omega_0 t} + \braket{\hat{b} \hat{b}^{\dag}} + \braket{\hat{b}^{\dag} \hat{b}} + \braket{\hat{b}^{\dag} \hat{b}^{\dag}} e^{2i\omega_0 t} \right), \\
&\braket{\hat{x}(0) \hat{x}(0)} = x_{\rm zpf}^2 \left( \braket{\hat{b} \hat{b}} + \braket{\hat{b} \hat{b}^{\dag}} + \braket{\hat{b}^{\dag} \hat{b}} + \braket{\hat{b}^{\dag} \hat{b}^{\dag}} \right).
\label{2pcorr}
\end{split}
\end{equation}
\noindent To determine the correlators of the ladder operators found in the above equation, we use Eqs.~(\ref{ACFdefqu}) and (\ref{tracedef}) along with the properties of the ladder operators given in Eqs. (\ref{raiselower}), (\ref{raiselowerh}) and (\ref{numops}) to find
\begin{equation}
\begin{split}
\braket{\hat{b}^{\dag} \hat{b}} &= \frac{ \displaystyle\sum\limits_{n} e^{-\beta E_n} \bra{n}\hat{b}^{\dag} \hat{b} \ket{n}}{\displaystyle\sum\limits_{n} e^{-\beta E_n}} = \frac{ \displaystyle\sum\limits_{n} e^{-\beta E_n} n \braket{n|n}}{\displaystyle\sum\limits_{n} e^{-\beta E_n}} = \frac{ \displaystyle\sum\limits_{n} n e^{-\beta E_n}}{\displaystyle\sum\limits_{n} e^{-\beta E_n}} = \braket{n}, \\
\braket{\hat{b} \hat{b}^{\dag}} &= \frac{ \displaystyle\sum\limits_{n} e^{-\beta E_n} \bra{n}\hat{b} \hat{b}^{\dag} \ket{n}}{\displaystyle\sum\limits_{n} e^{-\beta E_n}} = \frac{ \displaystyle\sum\limits_{n} e^{-\beta E_n} (n+1) \braket{n|n}}{\displaystyle\sum\limits_{n} e^{-\beta E_n}} = \frac{ \displaystyle\sum\limits_{n} (n+1) e^{-\beta E_n}}{\displaystyle\sum\limits_{n} e^{-\beta E_n}} = \braket{n} + 1, \\
\braket{\hat{b} \hat{b}} &= \frac{ \displaystyle\sum\limits_{n} e^{-\beta E_n} \bra{n}\hat{b} \hat{b} \ket{n}}{\displaystyle\sum\limits_{n} e^{-\beta E_n}} = \frac{ \displaystyle\sum\limits_{n} e^{-\beta E_n} \sqrt{n(n-1)} \braket{n|n-2}}{\displaystyle\sum\limits_{n} e^{-\beta E_n}} = 0, \\
\braket{\hat{b}^{\dag} \hat{b}^{\dag}} &= \frac{ \displaystyle\sum\limits_{n} e^{-\beta E_n} \bra{n}\hat{b}^{\dag} \hat{b}^{\dag} \ket{n}}{\displaystyle\sum\limits_{n} e^{-\beta E_n}} = \frac{ \displaystyle\sum\limits_{n} e^{-\beta E_n} \sqrt{n(n+1)} \braket{n|n+2}}{\displaystyle\sum\limits_{n} e^{-\beta E_n}} = 0. \\
\label{bcorrs}
\end{split}
\end{equation}
\noindent As expected, only the correlators with one creation and one annihilation operator are nonzero. We have also introduced $\braket{n}$ which is the average thermal population of quanta and is given by
\begin{equation}
\braket{n} = \frac{\displaystyle\sum\limits_{n=0}^{\infty} n e^{-\beta \hbar \omega_0 (n + 1/2)}}{\displaystyle\sum\limits_{n=0}^{\infty}e^{-\beta  \hbar \omega_0 (n + 1/2)}} = \frac{\displaystyle\sum\limits_{n=0}^{\infty} n e^{-\beta \hbar \omega_0 n}}{\displaystyle\sum\limits_{n=0}^{\infty}e^{-\beta  \hbar \omega_0 n}} = \frac{\frac{e^{-\beta  \hbar \omega_0}}{(1 - e^{-\beta  \hbar \omega_0})^2}}{\frac{1}{1 - e^{-\beta  \hbar \omega_0}}} = \frac{1}{e^{\beta \hbar \omega_0} - 1},
\label{navecalc}
\end{equation}
\noindent where in the above equation we have used Eq.~(\ref{QHOenergy}) to input an expression for $E_n$, along with the following sums \cite{gradshteyn}
\begin{equation}
\begin{split}
\displaystyle\sum\limits_{k=0}^{\infty} w^k &= \frac{1}{1-w}, \\
\displaystyle\sum\limits_{k=0}^{\infty} k w^k &= \frac{w}{(1-w)^2}.
\label{geomsums}
\end{split}
\end{equation}
\noindent It should be noted that the value we obtained for $\braket{n}$, known as the Bose-Einstein occupation factor, is exactly what we would expect for the average occupation number for a thermal distribution of bosons. 

Using the relations given in (\ref{bcorrs}), we can then express our first-order position correlators as
\begin{equation}
\begin{split}
&\braket{\hat{x}(t) \hat{x}(0)} = x_{\rm zpf}^2 \left[ \left( \braket{n} + 1 \right) e^{-i\omega_0 t} + \braket{n} e^{i\omega_0 t}  \right], \\
&\braket{\hat{x}(t) \hat{x}(t)} = \braket{\hat{x}(0) \hat{x}(0)} = x_{\rm zpf}^2 \left[ 2 \braket{n} + 1 \right].
\label{2pcorrres}
\end{split}
\end{equation}
\noindent These correlators are the ones given in Eq.~(\ref{ACFk2pcorr}), which are used to determine the $k$th-order quantum ACF.

\section{Contour Integration of the First-Order PSD}
\label{PSD1contint}

Here we use contour integration to calculate the integral given in Eq.~(\ref{x2avedefcl}). We begin by factoring the denominator of Eq.~(\ref{PSD1defcl}) in terms of its complex zeros using the quadratic equation, allowing us to write
\begin{equation}
\begin{split}
\bar{S}_{xx}(\omega) &= \frac{\bar{S}^{\rm th}_{FF}}{m^2 \left[\omega - \frac{1}{2} \left( i\Gamma + \sqrt{4 \omega_0^2 - \Gamma^2} \right) \right] \left[\omega - \frac{1}{2} \left( i\Gamma - \sqrt{4 \omega_0^2 - \Gamma^2} \right) \right]} \\ &\times \frac{1}{\left[\omega + \frac{1}{2} \left( i\Gamma - \sqrt{4 \omega_0^2 - \Gamma^2} \right) \right] \left[\omega + \frac{1}{2} \left( i\Gamma + \sqrt{4 \omega_0^2 - \Gamma^2} \right) \right]}.
\label{PSD1clpoles}
\end{split}
\end{equation}
\noindent We now have our linear displacement PSD in a form similar to that given in Eq.~(\ref{polefunc}), allowing us to easily compute the residues corresponding to the different poles. Inspecting the above equation, we can read off these poles as $\omega = \pm \frac{1}{2} \left(i \Gamma \pm \sqrt{4 \omega_0^2 - \Gamma^2} \right)$. Since we are in the high-$Q$ limit, we are assured that the square root quantities will be real, as $4 \omega_0^2 > \Gamma^2$. Therefore, only two of these poles will be in the top half of the complex plane, namely $\omega_1 = \frac{1}{2} \left( i \Gamma + \sqrt{4 \omega_0^2 - \Gamma^2} \right)$ and $\omega_2 = \frac{1}{2} \left( i \Gamma - \sqrt{4 \omega_0^2 - \Gamma^2} \right)$. We can then write the integral in Eq.~(\ref{x2avedefcl}) as
\begin{equation}
\left< x^2 \right> = \frac{1}{2 \pi}\int_{-\infty}^{\infty} \! \bar{S}_{xx}(\omega) \, d \omega = i \left[ {\rm Res}(\bar{S}_{xx},\omega_1) + {\rm Res}(\bar{S}_{xx},\omega_2)  \right].
\label{x2contint}
\end{equation}
\noindent Computing each of these residues separately using Eq.~(\ref{resdef}), we find
\begin{equation}
\begin{split}
{\rm Res}(\bar{S}_{xx}, \omega_1) = \frac{\bar{S}^{\rm th}_{FF}}{im^2 \Gamma \sqrt{4\omega_0^2 - \Gamma^2} \left( i \Gamma + \sqrt{4\omega_0^2 - \Gamma^2} \right)  }, \\
{\rm Res}(\bar{S}_{xx}, \omega_2) = \frac{-\bar{S}^{\rm th}_{FF}}{im^2 \Gamma \sqrt{4\omega_0^2 - \Gamma^2} \left( i \Gamma - \sqrt{4\omega_0^2 - \Gamma^2} \right)  }.
\label{PSD1res}
\end{split}
\end{equation}
\noindent Putting these results into Eq.~(\ref{x2contint}), we obtain
\begin{equation}
\left< x^2 \right> = \frac{\bar{S}^{\rm th}_{FF}}{2m^2 \omega_0^2 \Gamma}.
\label{x2resapp}
\end{equation}
\noindent Combining this result with Eq.~(\ref{x2averescl}), we can now determine $\bar{S}^{\rm th}_{FF}$, allowing us to properly normalize the first-order classical PSD.

\section{Contour Integration to Determine the Functional Form of the Second-Order PSD}
\label{contintchi}

In order to determine the second-order displacement PSD, we need to perform the convolution integral $\chi (\omega) * \chi (\omega )$. To evaluate this integral we follow a strategy similar to the previous section, in which we express the convolution integral given in Eq.~(\ref{chiconv}) as
\begin{equation}
\chi (\omega) * \chi (\omega) = \int_{-\infty}^{\infty} \! \chi(\omega') \chi(\omega - \omega') \, d \omega' =  \frac{1}{m^2} \int_{-\infty}^{\infty} \! y(\omega,\omega') \, d \omega',
\label{convchiapp}
\end{equation}
\noindent where $y(\omega,\omega')$ is a function expressing the integral in terms of its poles and is given by
\begin{equation}
\begin{split}
&y(\omega,\omega') = \frac{1}{\left[ \omega' + \frac{1}{2} \left( i \Gamma + \sqrt{4\omega_0^2 - \Gamma^2} \right) \right]\left[ \omega' + \frac{1}{2} \left( i \Gamma - \sqrt{4\omega_0^2 - \Gamma^2} \right) \right]} \\ &\times \frac{1}{\left[ \omega' - \omega - \frac{1}{2} \left( i \Gamma + \sqrt{4\omega_0^2 - \Gamma^2} \right) \right]\left[ \omega' - \omega - \frac{1}{2} \left( i \Gamma - \sqrt{4\omega_0^2 - \Gamma^2} \right) \right]}.
\label{ypole}
\end{split}
\end{equation}
\noindent In this form, the poles are easily read off at $\omega' = - \frac{1}{2} \left(i \Gamma \pm \sqrt{4\omega_0^2 - \Gamma^2} \right)$ and $\omega' = \omega + \frac{1}{2} \left(i \Gamma \pm \sqrt{4\omega_0^2 - \Gamma^2} \right)$. Again, choosing the semicircle that covers the top half of the complex plane, we concern ourselves with the two positive poles enclosed in this region, namely $\omega'_1 = \omega + \frac{1}{2} \left(i \Gamma + \sqrt{4\omega_0^2 - \Gamma^2} \right)$ and $\omega'_2 = \omega + \frac{1}{2} \left(i \Gamma - \sqrt{4\omega_0^2 - \Gamma^2} \right)$. We can then express our convolution integral as
\begin{equation}
\chi (\omega) * \chi (\omega) =  \frac{2 \pi i}{m^2} \left[ {\rm Res}(y,\omega'_1) + {\rm Res}(y,\omega'_2)  \right].
\label{chiconvcontint}
\end{equation}
\noindent We calculate these residues separately, obtaining
\begin{equation}
\begin{split}
{\rm Res}(y, \omega'_1) = \frac{1}{ \sqrt{4\omega_0^2 - \Gamma^2} \left( \omega + i \Gamma + \sqrt{4\omega_0^2 - \Gamma^2} \right) \left( \omega + i \Gamma \right)}, \\
{\rm Res}(y, \omega'_2) = \frac{-1}{ \sqrt{4\omega_0^2 - \Gamma^2} \left( \omega + i \Gamma - \sqrt{4\omega_0^2 - \Gamma^2} \right) \left( \omega + i \Gamma \right)}.
\label{convchires}
\end{split}
\end{equation}
\noindent Inputting these two residues in Eq.~(\ref{chiconvcontint}), we get
\begin{equation}
\chi (\omega) * \chi (\omega) = \frac{-4 \pi i}{m^2  \left( \omega + i \Gamma \right) \left( \omega^2 - 4 \omega_0^2 + 2i\omega \Gamma \right)}.
\label{chiconvresapp}
\end{equation}
\noindent This is the expression that allows for determination of the functional form of the second-order PSD for the classical damped harmonic oscillator.

\section{Contour Integration of the Second-Order PSD}
\label{PSD2contint}

We use contour integration one last time in order to determine the normalization constant $\bar{S}^{\rm th}_{F^2 F^2}$ for the classical second-order PSD. Following the same procedure as in the last two sections, we begin by writing the second-order PSD given in Eq.~(\ref{PSD2unnormcl}) in terms of its poles 
\begin{equation}
\begin{split}
&\bar{S}_{x^2 x^2}(\omega) = \frac{16 \pi^2 \bar{S}_{F^2 F^2}^{\rm th}}{m^4  \left( \omega + i \Gamma \right) \left( \omega - i \Gamma \right) \left( \omega + i \Gamma + \sqrt{4\omega_0^2 -\Gamma^2} \right)} \\ &\times \frac{1}{ \left( \omega + i \Gamma - \sqrt{4\omega_0^2 -\Gamma^2} \right) \left( \omega - i \Gamma + \sqrt{4\omega_0^2 -\Gamma^2} \right) \left( \omega - i \Gamma - \sqrt{4\omega_0^2 -\Gamma^2} \right)}.
\label{PSD2poles}
\end{split}
\end{equation}
\noindent Upon inspection of this equation, we can see that the poles are given by $\omega = \pm i \Gamma$ and $\omega = \pm i \Gamma \pm \sqrt{4 \omega_0^2 + \Gamma^2}$. We now look to compute the integral given in Eq.~(\ref{x4avedefcl}). Using our usual method, we concern ourselves with the three poles in the upper half of the complex plane, given by $\omega_1 = i \Gamma$, $\omega_2 = i \Gamma - \sqrt{4 \omega_0^2 - \Gamma^2}$ and $\omega_3 = i \Gamma + \sqrt{4 \omega_0^2 - \Gamma^2}$. We can then express the integral in question as
\begin{equation}
\left< x^4 \right> = \frac{1}{2 \pi}\int_{-\infty}^{\infty} \! \bar{S}_{x^2 x^2}(\omega) \, d \omega = i \left[ {\rm Res}(\bar{S}_{x^2 x^2},\omega_1) + {\rm Res}(\bar{S}_{x^2 x^2},\omega_2) + {\rm Res}(\bar{S}_{x^2 x^2},\omega_3) \right].
\label{x4contint}
\end{equation}
\noindent Computing each of the residues separately we find
\begin{equation}
\begin{split}
{\rm Res}(\bar{S}_{x^2 x^2},\omega_1) &= \frac{-8 \pi^2 \bar{S}_{F^2 F^2}^{\rm th}}{i \Gamma m^4 \left( 4 \omega_0^2 - \Gamma^2 \right) \left( 2 i \Gamma + \sqrt{4 \omega_0^2 - \Gamma^2} \right) \left( 2 i \Gamma - \sqrt{4 \omega_0^2 - \Gamma^2} \right) }, \\
{\rm Res}(\bar{S}_{x^2 x^2},\omega_2) &= \frac{2 \pi^2 \bar{S}_{F^2 F^2}^{\rm th}}{i \Gamma m^4 \left( 4\omega_0^2 - \Gamma^2 \right) \left( 2 i \Gamma - \sqrt{4 \omega_0^2 - \Gamma^2} \right) \left( i \Gamma - \sqrt{4 \omega_0^2 - \Gamma^2} \right)}, \\
{\rm Res}(\bar{S}_{x^2 x^2},\omega_3) &= \frac{2 \pi^2 \bar{S}_{F^2 F^2}^{\rm th}}{i \Gamma m^4 \left( 4\omega_0^2 - \Gamma^2 \right) \left( 2 i \Gamma + \sqrt{4 \omega_0^2 - \Gamma^2} \right) \left( i \Gamma + \sqrt{4 \omega_0^2 - \Gamma^2} \right)}.
\label{PSD2res}
\end{split}
\end{equation}
\noindent Inputting these residues into Eq.~(\ref{x4contint}), we obtain
\begin{equation}
\left< x^4 \right> = \frac{3 \pi^2 \bar{S}_{F^2 F^2}^{\rm th}}{\Gamma \omega_0^2  m^4 \left( 4 \omega_0^2 + 3 \Gamma^2\right) },
\label{x4resapp}
\end{equation}
\noindent from which we can determine $\bar{S}_{F^2 F^2}^{\rm th}$ using Eq.~(\ref{x4avedefcl}), allowing us to normalize the second-order classical PSD.

\section{Calculation of the First-Order Damped Quantum PSD using Input-Output Theory}
\label{PSDdamp}

In this section, we use the expressions for a harmonic oscillator coupled to an external heat bath given in Section \ref{QHO} to calculate the first-order PSD for a damped harmonic oscillator. Inputting Eq.~(\ref{xofw}) into Eq.~(\ref{PSD1dampdefqu}), we can write our damped PSD as
\begin{equation}
\begin{split}
\tilde{S}_{xx}(\omega) &= \frac{x_{\rm zpf}^2}{2 \pi}\int_{-\infty}^{\infty} \! \left[ \braket{\hat{b}_\gamma(\omega) \hat{b}^\dag_\gamma(\omega')} + \braket{\hat{b}^\dag_\gamma(\omega) \hat{b}_\gamma(\omega')} \right] \, d \omega'  \\ 
&= \frac{\Gamma x_{\rm zpf}^2}{2 \pi}\int_{-\infty}^{\infty} \! \bigg[ \frac{\braket{\hat{b}_n(\omega) \hat{b}^\dag_n(\omega')}}{\left(i(\omega_0 - \omega) + \Gamma/2 \right) \left(-i(\omega_0 + \omega') + \Gamma/2 \right)} + \frac{\braket{\hat{b}^\dag_n(\omega) \hat{b}_n(\omega')}}{\left(-i(\omega_0 + \omega) + \Gamma/2 \right) \left(i(\omega_0 - \omega') + \Gamma/2 \right)} \bigg] \, d \omega' ,
\label{PSD1dampbcorrs}
\end{split}
\end{equation}
\noindent where we have taken $\braket{\hat{b}_\gamma(\omega) \hat{b}_\gamma(\omega')} = \braket{\hat{b}^\dag_\gamma(\omega) \hat{b}^\dag_\gamma(\omega')} = 0$ due to the last line of Eq.~(\ref{noisecorrw}). Using the other two relations of Eq.~(\ref{noisecorrw}), we can perform the integrals in Eq.~(\ref{PSD1dampbcorrs}) to find
\begin{equation}
\tilde{S}_{xx}(\omega) = \Gamma x_{\rm zpf}^2 \left[ \frac{\braket{n}+1}{(\omega - \omega_0)^2 + (\Gamma/2)^2} + \frac{\braket{n}}{(\omega + \omega_0)^2 + (\Gamma/2)^2} \right].
\label{PSD1dampres}
\end{equation}
\noindent Here we have taken $n_b(\omega_0) = \braket{n}$ due to the fact that in thermal equilibrium the oscillator and heat bath will be at the same temperature.

\section{Classical Correspondence Approximations}
\label{ccapp}

In this section, we will look at the approximations made in order to show classical correspondence between the damped PSDs of the quantum harmonic oscillator and the classical harmonic oscillator. 

\subsection{First-Order}
\label{ccapp1}

We begin by using the approximations given in Eq.~(\ref{avenapprox}) to obtain the expression in the first line of Eq.~(\ref{PSD1ccapprox})
\begin{equation}
\begin{split}
\tilde{S}_{xx}(\omega) &\approx \frac{\Gamma k_B T}{2 m \omega_0^2}\left[ \frac{1}{(\omega - \omega_0)^2 + (\Gamma / 2)^2} + \frac{1}{(\omega + \omega_0)^2 + (\Gamma / 2)^2}\right] \\
&=\frac{\Gamma k_B T}{m \omega_0^2} \frac{\omega^2 + \omega_0^2 + (\Gamma / 2)^2}{ (\omega^2 - \omega_0^2)^2 + (\omega - \omega_0)^2 (\Gamma / 2)^2 + (\omega + \omega_0)^2 (\Gamma / 2)^2 + (\Gamma / 2)^4}.
\label{PSD1ccquapp}
\end{split}
\end{equation}
\noindent At this point, we apply the high-$Q$ approximation for which we neglect the last terms in both the numerator and denominator. As well, in this approximation the only significant contributions to the PSD occur at $\omega \approx \pm \omega_0$, otherwise the value will be small compared the the peak value. Inputting this approximation into our above PSD (we can use either sign, both give the same result since the classical PSD is an even function of $\omega$), we obtain
\begin{equation}
\begin{split}
\tilde{S}_{xx}(\omega) &\approx \frac{\Gamma k_B T}{m \omega_0^2} \frac{2 \omega_0^2}{ (\omega^2 - \omega_0^2)^2 + (2 \omega)^2 (\Gamma / 2)^2} \\
&= \frac{2 \Gamma k_B T}{m \left[ (\omega^2 - \omega_0^2)^2 + \omega^2 \Gamma^2 \right]},
\label{PSD1ccclapp}
\end{split}
\end{equation}
\noindent which is exactly the result we obtained in Eq.~(\ref{PSD1fincl}).

\subsection{Second-Order}
\label{ccapp2}

We follow a similar procedure here as in the previous section to show the classical correspondence of the second-order quantum PSD in the high-$Q$ limit. Beginning with the high temperature approximation of the second-order quantum PSD given by Eq.~(\ref{PSD2ccapprox}), we have
\begin{equation}
\begin{split}
\tilde{S}_{x^2 x^2}(\omega) &\approx \Gamma \left( \frac{k_B T}{m \omega_0^2} \right)^2 \left[ \frac{1}{\left( \omega - 2\omega_0 \right)^2 + \Gamma^2} + \frac{1}{\left( \omega + 2\omega_0 \right)^2 + \Gamma^2} + \frac{4}{\omega^2 + \Gamma^2} \right]\\
&= \Gamma \left( \frac{k_B T}{m \omega_0^2} \right)^2  \bigg[ \left(\omega^2 + \Gamma^2 \right) \left( \left( \omega + 2 \omega_0^2 \right)^2 + \left( \omega - 2 \omega_0^2 \right)^2 + 2 \Gamma^2 \right) + \left( \left( \omega + 2 \omega_0^2 \right)^2 + \Gamma^2 \right)\left( \left( \omega - 2 \omega_0^2 \right)^2 + \Gamma^2 \right) \bigg] 
\\ &\times \frac{1}{\left( \omega^2 + \Gamma^2 \right) \left( \left(\omega^2 - 4 \omega_0^2 \right)^2 + \Gamma^2 \left( \omega - 2 \omega_0 \right)^2 + \Gamma^2 \left( \omega + 2 \omega_0 \right)^2 + \Gamma^4  \right)}.
\label{PSD2ccquapp}
\end{split}
\end{equation}
\noindent As before, due to our high-$Q$ approximation, we only concern ourselves with the PSD in the vicinity of the peaks, namely $\omega \approx \pm 2 \omega_0$ and $\omega \approx 0$. As well, we neglect all terms of order $\Gamma^2$ or higher in the numerator and the $\Gamma^4$ term in the denominator. With these approximations we find
\begin{equation}
\begin{split}
\tilde{S}_{x^2 x^2}(\omega) &\approx \Gamma \left( \frac{k_B T}{m \omega_0^2} \right)^2  \frac{64 \omega_0^4}{\left( \omega^2 + \Gamma^2 \right) \left( \left(\omega^2 - 4 \omega_0^2 \right)^2 + 4 \Gamma^2 \omega^2 \right)} \\
&= \frac{64 \Gamma \left( k_B T \right)^2}{m^2 \left( \omega^2 + \Gamma^2 \right) \left( \left( \omega^2 - 4 \omega_0^2 \right)^2 + 4 \omega^2 \Gamma^2 \right)},
\label{PSD2ccclapp}
\end{split}
\end{equation}
\noindent where again we have recovered the classical PSD that we found in Eq.~(\ref{PSD2highQcl}).

\section{Quantum Power Spectral Density Calculations Using the Fluctuation Dissipation Theorem}
\label{PSDFDT}

Here we calculate the first- and second-order PSDs for the quantum harmonic oscillator using the fluctuation-dissipation theorem, providing an independent check on our results found in Eqs.~(\ref{PSD1finqu}) and (\ref{PSD2finqu}).

\subsection{First-Order}
\label{PSD1FDT}

The first-order PSD for the position operator can also be derived using the finite-temperature Green's function formalism \cite{mahan}. The main building block in this formalism is the time-ordered Green's function for the bosonic operator $\hat{b}$,
\begin{equation}
G(\tau)=\langle \mathcal{T}_\tau \hat{b}(\tau)\hat{b}^\dag(0)\rangle,
\label{G1}
\end{equation}
\noindent from which correlation functions of arbitrary operators can be constructed. Here we have introduced $\mathcal{T}_\tau$, which is a time-ordering operator on the Matsubara contour. In particular, for the purpose of computing the first-order PSD for the position operator one first computes a time-ordered correlation function for $\hat{x}(\tau)$ \cite{mahan},
\begin{equation}
\Pi_{xx}(\tau)=\langle \mathcal{T}_\tau \hat{x}(\tau)\hat{x}(0)\rangle=x_{\rm zpf}\left(G(\tau)+G(-\tau)\right),
\label{TOcorr}
\end{equation}
\noindent where the imaginary time $\tau$ takes values between zero and the inverse temperature $\beta=1/k_BT$. Note that we have also used the fact that the order of bosonic operators can be rearranged at will within a time-ordered product, and that for a time-independent Hamiltonian the Green's function (\ref{G1}) depends only on the difference of its two time coordinates. Both the  time-ordered correlation functions and the Green's function of bosonic operators are periodic in imaginary time and can be given Fourier series expansions,
\begin{equation}
\begin{split}
G(\tau)&=\frac{1}{\beta}\sum_{i\omega_n}G(i\omega_n) e^{-i\omega_n\tau},\\
\Pi(\tau)&=\frac{1}{\beta}\sum_{i\omega_n}\Pi(i\omega_n) e^{-i\omega_n\tau},
\label{fourseries}
\end{split}
\end{equation}
\noindent where $\omega_n=2n\pi/\beta$, $n\in\mathbb{Z}$ are bosonic Matsubara frequencies \cite{mahan}, and the Fourier coefficients are given by
\begin{equation}
\begin{split}
G(i\omega_n)&=\int_0^\beta \! G(\tau) e^{i\omega_n\tau} \, d\tau, \\
\Pi(i\omega_n)&=\int_0^\beta \! \Pi(\tau) e^{i\omega_n\tau} \, d\tau.
\label{fourcoef}
\end{split}
\end{equation}
\noindent We point out that for a single bosonic mode at frequency $\omega_0$, $G(i\omega_n)$ is given by
\begin{equation}
G(i\omega_n)=\frac{1}{\omega_0 - i\omega_n}.
\label{G1pole}
\end{equation}
\noindent $S_{xx}(\omega)$ can then be determined using the fluctuation-dissipation theorem \cite{mahan}
\begin{equation}
S_{xx}(\omega)= 2(n_B(\omega)+1) {\rm Im} \{ \Pi^R_{xx}(\omega) \},
\label{PSD1FDTdef}
\end{equation}
\noindent where $n_B(\omega)=(e^{\beta \hbar \omega}-1)^{-1}$ is the Bose factor at frequency $\omega$ and we have introduced the retarded correlation function $\Pi^R_{xx}(\omega)$, defined as the analytic continuation to real frequencies of the Matsubara correlation function $\Pi_{xx}(i\omega_n)$,
\begin{equation}
\Pi^R_{xx}(\omega)=\lim_{i\omega_n\rightarrow\omega+i\eta}\Pi_{xx}(i\omega_n),
\label{TOcorrret1}
\end{equation}
\noindent where $\eta$ is a positive infinitesimal. To determine this retarded correlation function, we first calculate the time-ordered correlation function by putting Eq.~(\ref{G1pole}) into Eq.~(\ref{TOcorr}) to obtain
\begin{equation}
\Pi_{xx}(i\omega_n)=x_{\rm zpf}^2\left(\frac{1}{i\omega_n + \omega_0}-\frac{1}{i\omega_n-\omega_0}\right),
\label{TOcorrpoles}
\end{equation}
\noindent thus
\begin{equation}
\Pi^R_{xx}(\omega)=x_{\rm zpf}^2\left(\frac{1}{\omega+i\eta+\omega_0}-\frac{1}{\omega+i\eta-\omega_0}\right).
\label{TOcorrretpoles}
\end{equation}
\noindent Using the identity $\displaystyle \lim_{\eta\rightarrow 0^+}(y+i\eta)^{-1}=\mathcal{P}(1/y)-i\pi\delta(y)$ for $y$ real where $\mathcal{P}$ stands for the Cauchy principal value, along with the relation $n_B(-\omega)=-[n_B(\omega)+1]$, the fluctuation-dissipation theorem (\ref{PSD1FDTdef}) gives
\begin{equation}
S_{xx}(\omega)=2\pi x_{\rm zpf}^2\left[(\langle n\rangle+1)\delta(\omega-\omega_0) + \langle  n \rangle \delta(\omega + \omega_0) \right],
\label{PSD1FDTres}
\end{equation}
\noindent where we used $n_B(\omega_0)=\langle n\rangle$. This result is in agreement with what we obtained for the first-order PSD in Eq.~(\ref{PSD1finqu}).

\subsection{Second-Order}
\label{PSD2FDT}

The second-order PSD for the position operator can also be derived in a similar way. One begins by expressing the position operator in terms of creation $\hat{b}^\dag$ and annihilation $\hat{b}$ operators. In this way, $\hat{x}^2$ can be written as $\hat{x}^2=x_{\rm zpf}^2(\hat{P}+\hat{Q})$, where we define $\hat{P}=\hat{b}\hat{b}+\hat{b}^\dag\hat{b}^\dag$ and $\hat{Q}=2\hat{N}+1$ with $\hat{N}=\hat{b}^\dag\hat{b}$ the number operator. The operator $\hat{Q}$ is special in that it commutes with the Hamiltonian in Eq.~(\ref{QHOhamnum}), such that it is a conserved quantity that acquires no time-dependence in the Heisenberg picture, $\hat{Q}(t)=\hat{Q}$. This, in turn, implies that correlation functions of $\hat{Q}$ will contain a delta function at zero frequency. In terms of the operators $\hat{P}$ and $\hat{Q}$, the real-time correlation function for $\hat{x}^2(t)$ can be written as
\begin{equation}
\langle\hat{x}^2(t)\hat{x}^2(0)\rangle=x_{\rm zpf}^4\bigl(\langle \hat{P}(t)\hat{P}(0)\rangle+\langle\hat{P}(t)\hat{Q}\rangle+\langle\hat{Q}\hat{P}(0)\rangle +\langle\hat{Q}^2\rangle\bigr).
\label{ACF2FDTdef}
\end{equation}
\noindent The expectation value of $\hat{Q}^2$ is given by
\begin{equation}
\langle\hat{Q}^2\rangle=\langle(2\hat{N}+1)^2\rangle=8\langle n\rangle(\langle n\rangle+1)+1,
\label{Q2FDT}
\end{equation}
\noindent in agreement with the time-independent term in the last line of Eq.~(\ref{ACF2defqu}). The second and third terms in Eq.~(\ref{ACF2FDTdef}) are zero because the operator $\hat{P}$ does not conserve the phonon number $\hat{N}$ while the operator $\hat{Q}$ does. We are thus left with the task of evaluating the correlation function of $\hat{P}$ appearing in the expression
\begin{equation}
\langle\hat{x}^2(t)\hat{x}^2(0)\rangle=x_{\rm zpf}^4\bigl(\langle\hat{P}(t)\hat{P}(0)\rangle+8\langle n\rangle(\langle n\rangle+1)+1\bigr).
\label{ACF2FDTP}
\end{equation}

To calculate $\langle\hat{P}(t)\hat{P}(0)\rangle$ at finite temperature, one again defines a time-ordered correlation function,
\begin{equation}
\Pi_{PP}(\tau)=\langle \mathcal{T}_\tau\hat{P}(\tau)\hat{P}(0)\rangle.
\label{TOPcorr2}
\end{equation}
\noindent The advantage of defining a time-ordered correlation function is that for Hamiltonians quadratic in the creation and annihilation operators, as is the case here, one can use Wick's theorem \cite{mahan} to express $\Pi_{PP}(\tau)$ as a product of Green's functions (\ref{G1}). Because the Hamiltonian commutes with the phonon number operator $\hat{N}$, expectation values of products of four creation operators or four annihilation operators vanish. Using Wick's theorem, we are left with
\begin{equation}
\begin{split}
\Pi_{PP}(\tau)&=\langle \mathcal{T}_\tau\hat{b}(\tau)\hat{b}(\tau)\hat{b}^\dag(0)\hat{b}^\dag(0)\rangle+\langle \mathcal{T}_\tau\hat{b}^\dag(\tau)\hat{b}^\dag(\tau)\hat{b}(0)\hat{b}(0)\rangle\nonumber\\
&=2\left(G^2(\tau)+G^2(-\tau)\right),
\label{TOPcorr2G}
\end{split}
\end{equation}
\noindent where we have made the same assumptions used in determining Eq.~(\ref{TOcorr}). Substituting the expressions given in Eq.~(\ref{fourseries}) into Eq.~(\ref{TOPcorr2G}), we obtain
\begin{equation}
\Pi_{PP}(i\omega_n)=\frac{2}{\beta}\sum_{ip_n}\bigl(G(ip_n)G(i\omega_n-ip_n)+G(ip_n)G(-i\omega_n-ip_n)\bigr),
\label{TOPcorr2Gfour}
\end{equation}
\noindent where $p_n$ are also bosonic Matsubara frequencies. Sums over Matsubara frequencies are most conveniently carried out by making use of the spectral function $A(\omega)$, defined in terms of the Green's function as
\begin{equation}
G(i \omega_n)= \frac{1}{2 \pi} \int \! \frac{A(\omega)}{\omega - i\omega_n} \, d \omega.
\label{spectfuncdef}
\end{equation}
\noindent hence from Eq.~(\ref{G1pole}) we can read off the spectral function,
\begin{equation}
A(\omega)=2\pi\delta(\omega-\omega_0).
\label{spectfuncres}
\end{equation}
\noindent Substituting Eq.~(\ref{spectfuncdef}) into Eq.~(\ref{TOPcorr2Gfour}), we have
\begin{equation}
\Pi_{PP}(i\omega_n) =\frac{2}{\left( 2 \pi \right)^2} \int \hspace{-4pt} \int \! A(\omega)A(\omega') \frac{1}{\beta}\sum_{ip_n}\biggl(\frac{1}{(ip_n-\omega)(i\omega_n-ip_n-\omega')} +\frac{1}{(ip_n-\omega)(-i\omega_n-ip_n-\omega')}\biggr) \, d \omega \, d \omega'. 
\label{TOPcorr2polesfour1}
\end{equation}
\noindent Such sums can be performed using contour integration \cite{mahan}, observing that the Bose factor $n_B(\omega)$ has single poles at the Matsubara frequencies $\omega=ip_n$ with residue $1/\beta$. One finds
\begin{equation}
\begin{split}
\Pi_{PP}(i\omega_n)=\frac{2}{\left( 2 \pi \right)^2}\int \hspace{-4pt} \int \! &A(\omega)A(\omega')(n_B(\omega)+n_B(\omega')+1)  \\ &\times \left(\frac{1}{i\omega_n+\omega+\omega'}-\frac{1}{i\omega_n-\omega-\omega'}\right) \, d \omega \, d \omega',
\label{TOPcorr2polesfour2}
\end{split}
\end{equation}
\noindent which upon making use of Eq.~(\ref{spectfuncres}) gives
\begin{equation}
\Pi_{PP}(i\omega_n)=2(2 \braket{n}+1)\left(\frac{1}{i\omega_n+2\omega_0}-\frac{1}{i\omega_n-2\omega_0}\right).
\label{TOPcorr2poles}
\end{equation}
\noindent Using Eq.~(\ref{TOcorrret1}), the retarded correlation function will be given as
\begin{equation}
\Pi_{PP}^R(\omega)=2(2\langle n\rangle+1)\left(\frac{1}{\omega+i\eta+2\omega_0}-\frac{1}{\omega+i\eta-2\omega_0}\right),
\label{TOPcorrret2poles}
\end{equation}
\noindent Substituting this result in the fluctuation-dissipation theorem (\ref{PSD1FDTdef}) we arrive at the relation
\begin{equation}
S_{PP}(\omega)=4\pi(2\langle n\rangle+1)(n_B(\omega)+1) \bigl[\delta(\omega-2\omega_0)-\delta(\omega+2\omega_0)\bigr].
\label{SPPres}
\end{equation}
\noindent Using the identity
\begin{equation}
n_B(2\omega)=\frac{[n_B(\omega)]^2}{2n_B(\omega)+1},
\label{nB2om}
\end{equation}
\noindent we obtain
\begin{equation}
S_{PP}(\omega)=4\pi\left[(\langle n\rangle+1)^2\delta(\omega-2\omega_0)+\langle n\rangle^2\delta(\omega+2\omega_0)\right].
\label{SPPfin}
\end{equation}
\noindent Taking the Fourier transform of Eq.~(\ref{ACF2FDTP}) and using the definition in Eq.~(\ref{PSDdefqu}), the second-order quantum PSD is given as
\begin{equation}
S_{x^2 x^2}(\omega)=x_{\rm zpf}^4\left[S_{PP}(\omega)+2\pi(8\langle n\rangle(\langle n\rangle+1)+1)\delta(\omega)\right],
\label{PSD2resFDT}
\end{equation}
\noindent hence using Eq.~(\ref{SPPfin}) we find the second-order PSD for the position operator to be,
\begin{equation}
S_{x^2x^2}(\omega)=2\pi x_{\rm zpf}^4\Bigl[2(\langle n\rangle+1)^2\delta(\omega-2\omega_0)+2\langle n\rangle^2\delta(\omega+2\omega_0)+(8\langle n\rangle(\langle n\rangle+1)+1)\delta(\omega)\Bigr],
\label{PSD2finFDT}
\end{equation}
\noindent in agreement with Eq.~(\ref{PSD2finqu}).

\section*{Acknowledgements}

This work was supported by the University of Alberta, Faculty of Science; NSERC Canada; Alberta Innovates Technology Futures; the Canada Research Chair program; and the Canadian Institute for Advanced Research.

\end{widetext}

\bibliography{bibfile}

\end{document}